\newif\ifsubmode
\submodefalse

\newif\ifprintfig
\printfigtrue

\newif\ifemulate
\emulatetrue

\ifsubmode
  \documentstyle[12pt,aasms4,epsf]{article}
  \received{}
  \accepted{}
  \journalid{}{}
  \articleid{}{}
\else
   \ifemulate
     \documentstyle[emulateapj,epsf,graphicx]{article}
     \submitted{{\it Submitted for publication in ApJ}}
   \else
     \documentstyle[11pt,aaspp4,epsf]{article} 
     \slugcomment{{\it Submitted for publication in ApJ}}
   \fi
\fi

\lefthead{van den Bosch \& Dalcanton}
\righthead{Dark Matter versus Modified Newtonian Dynamics}

\newcommand{\etal}{{et al.~}}

\newcommand{\lta}{\lesssim}
\newcommand{\gta}{\gtrsim}

\newcommand{\lte}{\leq}

\newcommand{\kmsmpc}{\>{\rm km}\,{\rm s}^{-1}\,{\rm Mpc}^{-1}}
\newcommand{\kms}{\>{\rm km}\,{\rm s}^{-1}}
\newcommand{\mss}{\>{\rm m}\,{\rm s}^{-2}}

\newcommand{\Msun}{\>{\rm M_{\odot}}}
\newcommand{\Lsun}{\>{\rm L_{\odot}}}

\begin{document}

\title{Semi-Analytical Models for the Formation of Disk Galaxies -- \\
       II. Dark Matter versus Modified Newtonian Dynamics}

\ifemulate
  \author{Frank C. van den Bosch\altaffilmark{1} and Julianne J. Dalcanton}
  \affil{Department of Astronomy, University of Washington, Seattle, 
         WA 98195, USA  (vdbosch,jd)@astro.washington.edu}
\else
  \author{Frank C. van den Bosch\altaffilmark{1,2},
          Julianne J. Dalcanton\altaffilmark{3}}
  \affil{Department of Astronomy, University of Washington, Seattle, 
         WA 98195, USA}
\fi


\altaffiltext{1}{Hubble Fellow}

\ifemulate\else
  \altaffiltext{2}{{\tt vdbosch@astro.washington.edu}}
  \altaffiltext{3}{{\tt jd@astro.washington.edu}}
\fi


\ifsubmode\else
  \ifemulate\else
     \clearpage
  \fi
\fi


\ifsubmode\else
  \ifemulate\else
     \baselineskip=14pt
  \fi
\fi


\begin{abstract}
  We present detailed semi-analytical models for the formation of disk
  galaxies both in  a Universe dominated by  dark matter (DM), and  in
  one for which the force law  is given by modified Newtonian dynamics
  (MOND).  We tune  the  models  to  fit the observed    near-infrared
  Tully-Fisher (TF) relation,  and compare numerous predictions of the
  resulting  models  with observations.  The  DM   and MOND models are
  almost indistinguishable.  They  both yield  gas mass fractions  and
  dynamical mass-to-light ratios   which  are in  good  agreement with
  observations.  Both  models reproduce  the  narrow  relation between
  global  mass-to-light ratio   and central  surface  brightness,  and
  reveal a characteristic acceleration,  contrary to claims that these
  relations are  not predicted by DM  models.  Both models  require SN
  feedback in order to reproduce  the lack of high surface  brightness
  dwarf galaxies.   However, the introduction of  feedback to the MOND
  models  steepens the TF  relation and  increases the scatter, making
  MOND only marginally consistent with observations.  The most serious
  problem  for  the DM models  is their   prediction  of steep central
  rotation curves.  However, the  DM rotation curves are only slightly
  steeper  than those of   MOND, and are  only marginally inconsistent
  with the poor resolution data on LSB galaxies.
\end{abstract}



\ifemulate\else
   \clearpage
\fi


\section{Introduction}
\label{sec:intro}

The most successful cosmological model to date  consists of a universe
dominated  by  cold   dark  matter (CDM), possibly     with a non-zero
cosmological constant.   However, this paradigm faces  several serious
shortcomings.  First  is   the high central densities   predicted  for
virialized dark  halos.   Navarro, Frenk  \&  White  (1996, 1997) have
claimed  the existence of a  universal  density profile for virialized
halos, with $\rho_{\rm DM}  \propto r^{-1}$ at small radii  (hereafter
referred  to as the  NFW profile).   More  recently, higher resolution
simulations have argued that CDM  halos should have even steeper cusps
(Fukushige \& Makino  1997; Moore \etal 1998;  Moore  \etal 1999b; but
see Kravtsov \etal  1998).    Several studies have argued   that these
steep cusps are inconsistent with the observed rotation curves of dark
matter dominated systems, such   as dwarfs and low  surface brightness
galaxies  (Flores \& Primack 1994;  Moore  1994; Burkert 1995; Navarro
1998;  McGaugh \& de Blok 1998a;  Stil  1999).  Secondly, Klypin \etal
(1999) and Moore  \etal  (1999a) have indicated  an additional problem
for the CDM paradigm:   simulations predict that galaxies should  have
many more satellite  systems than observed.   Although the  reason for
this  large discrepancy   may   be related to  observational   biases,
feedback  from supernovae, or  an  ionizing background, it might  also
indicate that   the CDM power   spectrum has too   much power on small
scales.  Both these problems may (partially)  be solved with a mixture
of  hot and  cold dark matter.  However, the  dark matter scenario  is
certainly  starting to loose  its appealing character,   if we have to
rely on a  hybrid mix of baryons, cold  dark matter, hot dark  matter,
and  a   non-zero cosmological   constant  in   order to   explain the
observations.  The  problems  with fine-tuning  become   more and more
severe,  and  it might  therefore   be   worthwhile to  explore   some
alternatives to the dark matter theory.

Although     the flat  rotation   curves   of  galaxies are  generally
interpreted as evidence for the  existence of dark matter, they  could
also  result from our theory of  gravity (or  inertia) being in error,
such that the  dynamical   masses inferred from the  usual   Newtonian
equations  are overestimated.  Several  groups have tried to eliminate
the  need for  dark matter by  modifying the  theory of gravity (e.g.,
Milgrom  1983a; Sanders 1986;  Mannheim \& Kazanas  1989; Liboff 1992;
Moffat  \& Sokolov  1996).   So far, the   modified Newtonian dynamics
(MOND) proposed by Milgrom (1983a) has been remarkably successful, and
in this  paper we compare  several predictions  from MOND  with  those
based on dark matter (hereafter DM).

Disk galaxies are,  from  a  dynamical  point of view,   fairly simple
systems  in which the   discrepancies between  the  luminous and   the
inferred  dynamical   masses  (the  ``mass  discrepancies'') are least
ambiguous.  In addition, they have been extensively studied, with data
available  for systems that span  several  orders of magnitude in both
mass and  surface brightness.   They are  thus the  ideal test-beds to
compare predictions based on MOND and DM with observations in the hope
to discriminate between the two alternatives.

For both MOND and DM to be considered  viable theories, they should be
able to explain the following properties of disk galaxies:

\begin{enumerate}

\item The slope, scatter,  and zero-point of the Tully-Fisher relation
  (Tully \& Fisher 1977).  In particular, the  fact that both high and
  low surface  brightness disks (HSBs  and LSBs,  respectively) follow
  the same relation without  any significant offset  (Sprayberry \etal
  1995;  Zwaan \etal 1995;  Hoffman   \etal 1996; Tully \&   Verheijen
  1997).

\item The shape of  the observed rotation  curves. In  particular, the
  systematic variation of  these curves  with surface brightness   and
  luminosity (Rubin,  Thonnard  \&  Ford 1980;  Burstein   \etal 1982;
  Casertano \& van Gorkum 1991; Persic, Salucci \& Stel 1996).

\item The  increase  of mass discrepancies with  decreasing luminosity
  and  surface brightness  (e.g., Salucci \&    Frenk 1989; Persic  \&
  Salucci  1988,  1990; Kormendy   1990;  Broeils 1992;  Forbes  1992;
  Persic, Salucci \&  Stel 1996; McGaugh  \& de Blok 1998a).   This is
  directly linked to item~2.

\item The increase  of gas mass  fractions with decreasing  luminosity
  and  surface brightness (Gavazzi  1993; Gavazzi,  Pierini \& Boselli
  1996; McGaugh \& de Blok 1997).

\item  The sharp  cut-off   in the  distribution  of  central  surface
  brightnesses at  the bright end, close  to the value of  the Freeman
  (1970) law (Allen \& Shu 1979; McGaugh, Bothun \& Schombert 1995; de
  Jong 1996a; McGaugh \& de Blok 1998a).

\item  The presence of  a characteristic acceleration  in the rotation
  curves of disk galaxies (McGaugh 1998).

\end{enumerate}

In a  series of papers, McGaugh  \& de Blok (1998a,b;  hereafter MB98a
and  MB98b) have used several  of  these observational  facts to argue
against the DM hypothesis  and in favor of  MOND.  In particular, they
argue  that the DM hypothesis   faces severe fine-tuning problems.  In
this paper we revisit these issues using more sophisticated models for
the  formation   of  disk galaxies   under   both the  DM   and MOND
hypotheses.  We  show that, contrary to the  claims  made by MB98a and
MB98b,   the DM model    can  explain  virtually  all   aforementioned
observations with  a  minimal amount of  fine-tuning.  Furthermore, we
show  that MOND  needs  a  similar amount   of  fine-tuning, therewith
loosing its apparent advantage over the DM scenario.

This paper    is organized as   follows.   In  \S\ref{sec:diskform} we
discuss  the  models,  both   for the DM   and   the MOND models.   In
\S\ref{sec:tf} we   use the   near-infrared Tully-Fisher  relation  to
constrain  the free parameters  of the models.  In \S\ref{sec:comp} we
compare  the    models to   data on    gas  mass  fractions, dynamical
mass-to-light     ratios,    and   characteristic   accelerations.     
\S\ref{sec:rcs} compares the detailed  rotation curve shapes from  the
DM     and   MOND  models.    We      summarize  our  conclusions   in
\S\ref{sec:concl}.  Throughout this  paper we adopt a Hubble  constant
of $H_0 =  70 \, h_{70} \, \kmsmpc$  in all our models.  Any  quantity
that depends on the distance scale is written in terms of $h_{70}$.
 
\section{The formation of disk galaxies}
\label{sec:diskform}

\subsection{The dark matter hypothesis}
\label{sec:DMform}

In van den   Bosch (1999; hereafter   Paper~I) we presented   detailed
models for the formation of disk galaxies in the context of a universe
dominated by  CDM.  Here we briefly summarize   the ingredients of the
models, and refer the reader to Paper~I for details.

In  our  models,  disks  form by  the settling  of  baryonic matter in
virialized dark halos, which are described by the NFW density profile.
We assume  that while the baryons radiate  their binding  energy, they
conserve their  specific angular momentum, thus   settling into a disk
with a scale length that  is proportional to  the angular momentum and
size  of the dark  halo (see also   Fall \& Efstathiou 1980; Dalcanton
Spergel \& Summers 1997; Mo, Mao  \& White 1998;  van den Bosch 1998). 
The angular momentum,  $J$, is thought  to originate from cosmological
torques and can be   characterized by a dimensionless  spin parameter,
$\lambda$,   the distribution of  which is  well constrained from both
analytical  and numerical  studies (e.g.,  Barnes  \& Efstathiou 1987;
Warren \etal 1992). For each separate model galaxy, we randomly draw a
value for $\lambda$ from this distribution.  The disk is assumed to be
a thin  exponential  with   surface  density $\Sigma(R)  =    \Sigma_0
\exp(-R/R_d)$. Adiabatic contraction is taken into account to describe
how the density profile of the dark matter changes  due to the cooling
of the baryonic matter (Blumenthal \etal  1986; Flores \etal 1993).  A
global stability  criterion is used to  investigate  if the disks that
form  are   stable.   If unstable,   part of    the disk material   is
transformed into a bulge  component, until the disk reaches  stability
(van den Bosch 1998).

Once the  density distribution of the  baryonic material  is known, we
compute the   fraction of baryons   converted  into stars.   Kennicutt
(1989) has  shown that the star formation  rate (SFR) in disk galaxies
follows a Schmidt law (Schmidt 1959), but is abruptly suppressed below
a  given threshold density. The  critical density is given by Toomre's
(1964) stability criterion as:
\begin{equation} \label{Toomre}
\Sigma_{\rm crit}(R) = {\sigma_{\rm gas} \, \kappa(R) \over 3.36 \, G
\, Q} 
\end{equation} 
Here $Q$ is a dimensionless constant near unity, $\sigma_{\rm gas}$ is
the velocity  dispersion of  the  gas, and  $\kappa$ is  the  epicycle
frequency, which can be derived from the rotation curve.

Setting $\sigma_{\rm gas} = 6 \kms$ and $Q = 1.5$ these star formation
thresholds were found  to  coincide with the radii  where $\Sigma_{\rm
gas}  = \Sigma_{\rm crit}$.  Following  an earlier suggestion by Quirk
(1972),  we therefore consider the  fraction of disk mass with surface
densities  in  excess of $\Sigma_{\rm crit}$   to be eligible for star
formation.   If one  further  considers   the  empirical Schmidt   law
determined by Kennicutt (1998) for a large sample of disk galaxies, it
is found that over the typical lifetime of a  galaxy virtually all the
mass  which is  eligible for  star formation  does  in fact  turn into
stars.  This  is consistent with the  lack of major star  formation in
the  inter-arm regions in spirals  and yields gas  mass fractions that
are in good agreement with observations (Paper I).  In what follows we
therefore assume that  the entire bulge mass  and the entire disk mass
with $\Sigma(R)>\Sigma_{\rm  crit}$ are transformed into stars, giving
a total stellar mass of $M_*$.
 
The total   $K$-band  luminosity  of  the  disk-bulge system   is then
calculated using  $L_K = M_{*}/\Upsilon^{*}_K$, where $\Upsilon^{*}_K$
is the   stellar mass-to-light ratio.   Since $\Upsilon_K^{*}$ depends
only weakly  on the age  and metallicity of  a stellar population (see
discussion in   Paper~I), we assume  a constant  stellar mass-to-light
ratio of $\Upsilon_K^{*} = 0.4 \, h_{70} \, \Msun/\Lsun$.

With the mass in stars known,  the total amount  of energy produced by
SN can be calculated, once we make assumptions about the energy per SN
($E_{\rm SN}$)  and the number  of SN per solar  mass of  stars formed
($\eta_{\rm SN}$).  We then calculate the total  baryonic mass that is
prevented   from becoming  part of  the  disk/bulge  system  due to SN
feedback as
\begin{eqnarray} \label{hotmass} 
\lefteqn{M_{\rm hot} = 3.22 \, M_{*} \, \varepsilon_{\rm SN}^0 \times}
 \;\;\; \nonumber \\ 
& \left({\eta_{\rm  SN} \over 0.004 \Msun^{-1}}\right)
\left({E_{\rm SN} \over 10^{51} {\rm erg}}\right) 
\left({V_{200} \over 250 \kms}\right)^{\nu - 2}.
\end{eqnarray}
Here $V_{200}$ corresponds to the circular velocity of the halo at the
radius $r_{200}$,  inside  of which the  average  density equals  200
times  the critical density  for closure, and $\varepsilon_{\rm SN}^0$
and $\nu$  are two free parameters that  describe the  efficiency with
which the  SN  energy is  used to prevent   gas from  settling  in the
disk/bulge system (see Paper~I).     Throughout we adopt a  value   of
$\eta_{\rm SN} = 0.004 \Msun^{-1}$ which corresponds to a Scalo (1986)
initial mass function.  An iterative procedure is  used to compute the
masses in  the various phases  in  a self-consistent  manner, i.e., we
ensure that $M_{200} =  M_{\rm  DM} + M_{*} +   M_{\rm cold} +  M_{\rm
  hot}$,  with  $M_{200}$ the total mass   within $r_{200}$ (before SN
blow-out),  and $M_{\rm cold}$  the disk mass that  is not turned into
stars.

\subsection{The modified Newtonian dynamics hypothesis}
\label{sec:MONDform}

We now   construct similar models  but  under the  hypothesis of MOND,
which  assumes that the force law  changes from conventional Newtonian
form when the   acceleration of  a  test particle  is  smaller than  a
characteristic acceleration $a_0$  (which  is a universal   constant).
The   true gravitational  acceleration ${\bf   a}$  is related to  the
Newtonian acceleration ${\bf a_N}$ as
\begin{equation}
\label{accMOND} \mu (a/a_0) \, {\bf a} = {\bf a_N} 
\end{equation} 
($a  = \vert {\bf a} \vert$).   The interpolation function $\mu(x)$ is
not specific,  but is required  to have the asymptotic behavior $\mu(x
\gg   1) \rightarrow  1$  (the  Newtonian regime) and   $\mu(x  \ll 1)
\rightarrow x$ (the MOND  regime).   With the  rotation law  given  as
usual, $V_c^2(r)/r =  a$,  the circular velocity  of  a finite bounded
mass $M$ in the limit $a \rightarrow 0$ becomes
\begin{equation}
\label{velMOND} 
V_{\infty} = (G\, M \, a_0)^{1/4}, 
\end{equation} 
with $G$ the gravitational constant.  MOND thus predicts
asymptotically flat rotation curves.

First we need to  set $a_0$ and the  interpolation function $\mu(x)$.  
We follow   previous  studies of MOND,   and set  $a_0  =  1.12 \times
10^{-10} \, h_{70} \, \mss$ and adopt
\begin{equation}
\label{mufunc}
\mu(x) = {x \over \sqrt{1 + x^2}}.
\end{equation}
For  these choices,  MOND yields  good  fits to  the observed rotation
curves of disk galaxies (Kent 1987;  Milgrom 1988; Begeman, Broeils \&
Sanders 1991; Sanders 1996;  de Blok \&  McGaugh 1988).  With $\mu(x)$
given by equation~(\ref{mufunc}), the circular velocity as function of
radius becomes
\begin{equation}
\label{vcircmond}
V_c(r) = V_N(r) \left[ {1 \over 2} + \sqrt{{1 \over 4} + {a_0^2 \, r^2
      \over V_N^4(r)}} \, \right]^{1/4},
\end{equation}
where $V_N(r) = \sqrt{r\,  {\rm d}\Phi /  {\rm  d}r}$ is  the circular
velocity derived for pure Newtonian dynamics.

Unfortunately, there is currently no MOND equivalent  to the theory of
structure formation in a  DM universe.  In  particular, we don't  know
the  formation epochs or  distribution  of angular momenta of galaxies
that form   in  a MOND   universe.  We  therefore follow an  empirical
procedure in which we assume that galaxy formation under MOND produces
disk galaxies with  the masses and  dimensions as observed.  As in the
DM scenario, disks are  assumed to be  thin exponentials.  We randomly
draw a total (baryonic) mass from the interval $10^8 h_{70}^{-1} \Msun
\lte M_{\rm tot}  \lte  5.7  \times 10^{11} h_{70}^{-1}   \Msun$.   We
initially set the   disk  mass, $M_d$,  equal  to  $M_{\rm tot}$,  and
randomly draw a central disk surface density from  the interval $45 \,
h_{70}  \, \Msun  {\rm pc}^{-2} \lte  \Sigma_0  \lte 2400 \, h_{70} \,
\Msun {\rm pc}^{-2}$. The scale length of the disk is then $R_d = (M_d
/ 2 \, \pi \, \Sigma_0)^{1/2}$.

The first step is  to investigate whether  the disk is  stable against
global  instabilities.  To that   extent  we use  the   same stability
criterion as for the DM scenario, i.e., a disk is considered stable if
\begin{equation}
\label{stabalpha}
\alpha = 0.3 {\sqrt{G \, M_d / R_d} \over \, V_c(3 R_d)} 
       < \alpha_{\rm crit},
\end{equation}
where $V_c(3  R_d)$ is the circular   velocity of the  galaxy at three
scale  lengths   (cf.  Christodolou,   Shloshman   \&   Tohline  1995;
Efstathiou, Lake \& Negroponte 1982; van den Bosch 1998).  In the case
of Newtonian dynamics  with no DM halos, $\alpha  = 0.5$.  However, in
the case of MOND, $\alpha <  0.5$, at least if  the acceleration at $3
\, R_d$ is  sufficiently small;  i.e., the  MOND  force  law helps  to
stabilize  disks (Milgrom 1989).   Unlike the  DM  scenario, for which
numerical simulations have shown  that $\alpha_{\rm crit} \simeq 0.35$
(Christodolou \etal 1995), no equivalent study exists based upon MOND.
We therefore consider $\alpha_{\rm crit}$ a free parameter.

Once  the density distribution of the  gas is known,  we calculate the
mass in  stars,  in the same  fashion  as for  the dark  matter models
discussed   above.   Note, however, that  in   the  calculation of the
epicycle frequency, $\kappa$, we  now use the circular  velocity based
on  the  modified   force  law.   The  $K$-band   luminosity  is again
calculated from   $M_{*}$  adopting  a  constant  mass-to-light  ratio
$\Upsilon_K^{*}$, which we consider a free parameter.

Finally, we include SN feedback, using the same  approach as above for
DM.   The   hot  gas  mass,    $M_{\rm   hot}$  is   calculated  using
equation~(\ref{hotmass}), but with  $V_{200}$ replaced by $V_{\infty}$
(equation~[\ref{velMOND}]).  We use an  iterative  technique to ensure
self-consistency, i.e., $M_{\rm  tot} = M_{*}  + M_{\rm cold} + M_{\rm
  hot}$.
 
\section{Constraints from the Tully-Fisher relation}
\label{sec:tf}

The  slope, scatter and zero-point  of the Tully-Fisher (hereafter TF)
relation contain information about the structure and formation of disk
galaxies. Here we use the observed characteristics of this fundamental
scaling relation  to set the  parameters of  our disk formation models
described in the previous section.

\subsection{Observations and simple dynamical predictions}
\label{sec:tfobs}

The slope and scatter of the empirical  TF relation depend strongly on
the luminosity and velocity  measures used, with the relation becoming
shallower going towards  bluer passbands  (Aaronson, Huchra \&   Mould
1979; Visvanathan 1981; Tully,    Mould \& Aaronson 1982; Wyse   1982;
Pierce \& Tully 1988;  Gavazzi 1993; Verheijen 1997).  Therefore, when
trying to infer constraints   on the structure  and formation  of disk
galaxies  from  the TF relation,  it  is absolutely essential that one
extracts the same luminosity   and velocity measures from   the models
(see discussions   in Courteau  1997, Verheijen   1997,  and Paper~I). 

The empirical  TF relation  which most directly  reflects the  mass in
stars and  the total  dynamical mass of   the halo is  the $K$-band TF
relation of Verheijen (1997), which uses the flat part of the rotation
curve as a characteristic velocity:
\begin{equation}
\label{TFfund}
L_K = 2.96 \times 10^{11} \, h_{70}^{-2}  \Lsun \, \left({V_{\rm flat}
  \over 250 \kms}\right)^{4.2}.
\end{equation}
The slope of this  TF relation is $b=-10.5  \pm 0.5$, and the observed
scatter is $\sigma_M = 0.29$ mag.  This empirical TF relation is shown
in the  upper left    panel  of Figure~\ref{fig:tf}.    Open   circles
correspond to the    galaxies in Verheijen's   sample  of  Ursa  Major
spirals,  and  the thick solid  line is  the  best-fitting TF relation
(equation~[\ref{TFfund}]).

Within the DM framework, simple dynamics  predict a TF relation of the
form
\begin{equation}
\label{tfdm}
L_K = {\epsilon_{\rm gf} \, f_{\rm bar} \over 10 \, G \, H_0 \,
  \Upsilon_K} \, \left({V_{200} \over V_{\rm flat}}\right)^3 \, 
V_{\rm flat}^3,
\end{equation}
(Dalcanton  \etal 1997;  White  1997; Mo \etal  1998;  Syer, Mao \& Mo
1999;  van den Bosch 1998;  Paper~I).  Here $\epsilon_{\rm gf}$ is the
galaxy formation  efficiency,   which describes what  fraction  of the
baryonic mass inside the DM halo ultimately  ends up in the disk/bulge
system, $f_{\rm bar}$  is the baryonic mass  fraction  of the Universe
($f_{\rm bar} =  \Omega_{\rm bar}/\Omega_0$), and $\Upsilon_K = M_{\rm
  gal}/L_K  =  \Upsilon_K^{*}    M_{\rm gal}/M_{*}$ is   the  $K$-band
mass-to-light ratio of the galaxy (disk plus bulge).

From equation~(\ref{velMOND}) it  is immediately apparent that  in the
case of MOND, the TF relation becomes
\begin{equation}
\label{tfmond}
L_K = {1 \over a_0 \, G \, \Upsilon_K} \, 
\left({V_{\infty} \over V_{\rm flat}}\right)^4 \, V_{\rm flat}^4,
\end{equation}
(see e.g., Milgrom 1983b; MB98b). 

Thus whereas DM  predicts a TF relation  with a slope  of $b =  -7.5$,
MOND  predicts $b=-10$, in  agreement  with the empirical relation  of
equation~(\ref{TFfund}). In the case of DM, extra physics are required
to steepen the predicted TF relation to the observed slope. MB98a have
presented a detailed discussion in which they  argue that this results
in a  serious  fine-tuning problem for  DM.   On  the other hand,  the
almost perfect   agreement between the    observed TF slope   and  the
pure-gravitational value for MOND (equation~[\ref{tfmond}])   presents
an anti-fine-tuning problem.  There is very little  room for any other
galaxy  characteristic   to vary  systematically   with  mass, without
changing the TF slope.

The above suggests that  simple  dynamical arguments  can be  used  to
predict  the existence of  an underlying TF  relation  for both DM and
MOND.   However, to   accurately predict  the  slope,  zero-point, and
scatter of the  TF relationship, it  is  necessary to incorporate  the
more detailed physics discussed  in \S\ref{sec:diskform}. We do so  in
the   following   two sections   for   the  DM  and MOND   hypotheses,
respectively. As mentioned in  \S\ref{sec:tfobs}  it is essential   to
extract the  same velocity measure from the  models as the one used in
the  data, and we   therefore extract $V_{\rm   flat}$ from  our model
galaxies (see paper~I for details).

\subsection{The Tully-Fisher relation and dark matter}
\label{sec:tfdark}

In Paper~I we showed  that taking the stability-derived star formation
threshold  densities into account     (with $Q=1.5$) steepens  the  TF
relation from   $L_K \propto V_{\rm   flat}^3$  (as predicted  by pure
dynamics) to $L_K \propto   V_{\rm  flat}^{3.6}$.  We argued that   to
further steepen the  TF relation to  its observed slope, feedback from
SN is required.  Fine-tuning   the  two parameters that  describe  the
efficiency with which  SN energy is used to  prevent gas from settling
in the disk/bulge system ($\varepsilon_{\rm SN}^0$ and $\nu$), we were
able to obtain a  TF relation with a  slope, zero-point, and an amount
of  scatter  that are all   in excellent agreement with  the empirical
relation  of  equation~(\ref{TFfund}).  This model  is  referred to as
model L5 in Paper~I, and we adopt the same convention here. This model
is based  on a cosmology   with $\Omega_0 = 0.3$,  $\Omega_{\Lambda} =
0.7$, $h_{70}  =  1$  and $\sigma_8 =   1.0$ (see  paper~I),  and  its
parameters are listed in  Table~\ref{tab:param}.  Model L5 predicts an
intrinsic TF scatter of  $\sigma_M \simeq 0.2$  mag, with no offset in
normalization between  HSB and   LSB  galaxies, as observed.   The  TF
relation  of model    L5 is  shown  in  the   upper   right  panel  of
Figure~\ref{fig:tf}.

\subsection{The Tully-Fisher relation and MOND}
\label{sec:tfmond}

Although simple  dynamics predict a TF relation  for MOND with a slope
$b=-10$, consistent  with observations,   this ignores the   fact that
$\Upsilon_K$ varies with    mass;  gas mass fractions  are    known to
increase with  decreasing  mass (see  \S\ref{sec:gasfrac}), and taking
this into account steepens the TF relation. It therefore remains to be
seen whether MOND does not actually predict  TF relations that are too
steep.   We  can investigate   this   using the  models described   in
\S\ref{sec:MONDform}.

As with the DM models, we set $Q=1.5$.  The stellar mass-to-light ratio
in the $K$-band, $\Upsilon_K^{*}$, is set by fitting the normalization
of the TF   relation, yielding $\Upsilon_K^{*} =   0.53 \, h_{70}   \,
\Msun/\Lsun$.    This value  is   consistent with  estimates based  on
stellar population models, which in itself can be considered a success
of MOND.

To set the free parameter  $\alpha_{\rm crit}$, with
the help  of   equation~(\ref{vcircmond})   the  stability   criterion
(equation~[\ref{stabalpha}]) can be written as
\begin{eqnarray}
\label{maxsb}
\lefteqn{\Sigma_0 \lte   1.91 \times 10^3 \, \left({a_0 \over 10^{-10}
      \mss}\right) \times } 
\;\;\;\;\;\;\;\;\;\;\;\;\;\;\;\;\;\;\;\;\;\;
\nonumber \\
& \left[\left(   {1 \over 8 \,  \alpha_{\rm
        crit}^4} - 1\right)^2 - 1\right]^{-1/2} \; \Msun {\rm pc}^{-2}.
\end{eqnarray}
The stability parameter $\alpha_{\rm  crit}$ thus  directly translates
into a maximum  surface  density of disks.  From   the  fact that  the
brightest  observed   disks have  central   surface   brightnesses  of
$\mu_{0,K} \simeq 16.5$ mag arcsec$^{-2}$  (e.g., de Jong 1996b; Tully
\etal    1996) we    infer   $\alpha_{\rm   crit}   \gta  0.45$   (for
$\Upsilon_K^{*} = 0.53 \, h_{70} \, \Msun/\Lsun$).  If we were to take
$\alpha_{\rm  crit} = 0.35$  (as for  the  DM models),  no disks  with
central  surface brightnesses  brighter than   $\mu_{0,K} = 18.1$  mag
arcsec$^{-2}$   would   be   allowed,   clearly      inconsistent with
observations.   Since in general not all  the  gas mass is transformed
into stars, the  limit of $\alpha_{\rm crit}  = 0.45$ is conservative. 
We find that  with $Q=1.5$  and $\Upsilon_K^{*} =  0.53 \,  h_{70}  \,
\Msun/\Lsun$ a value  of $\alpha_{\rm crit}  = 0.48$ yields $\mu_{0,K}
\lta  16.5$  mag arcsec$^{-2}$, and  we adopt  this  value throughout. 
None  of the results presented  here, however,  depend on this choice:
$\alpha_{\rm crit}$ merely   sets the maximum  central surface density
for disks with a  given bulge-to-disk ratio, and is  set such  that we
obtain  disks with the  full range of  observed values of $\mu_{0,K}$.
In mimicking the sample selection of Verheijen (1997) we only consider
model galaxies with  a bulge-to-disk ratio   $B/D \lte 0.2$  (see also
Paper~I).

\placefigure{fig:tf}

We compare  the DM model (L5) with  two different  MOND models, M1 and
M2.   The  former neglects SN feedback,  whereas  the latter considers
blow-out by  SN with $\varepsilon_{\rm SN}^0  = 0.05$ and $\nu = -3.0$
(see Table~\ref{tab:param}).   The   motivation  for  these particular
feedback  parameters will   be discussed in  \S\ref{sec:ups}.   The TF
relations for models  M1  and M2 are  shown   in the lower  panels  of
Figure~\ref{fig:tf}.  The star formation  recipe with $Q=1.5$ steepens
the  MOND    TF    relation  from   $L_K    \propto  V_{\rm   flat}^4$
(equation~[\ref{tfmond}])  to $L_K  \propto V_{\rm flat}^{4.2}$ (model
M1),   and brings  it  in  excellent agreement  with  the empirical TF
relation of Verheijen. The predicted,  intrinsic amount of scatter  is
only $\sigma_M =  0.21$ mag, once  again  in excellent  agreement with
observations.     The introduction  of  SN  feedback   steepens the TF
relation further, to $b=-11.9$ for model M2, and increases the scatter
to $\sigma_M = 0.33$ mag.  Model M2 thus yields  a TF relation that is
too steep.  In   addition, the amount of  scatter  is only  marginally
consistent with the data,   i.e., Verheijen (1997)   derived $\sigma_M
\lte 0.3$ mag at 95 percent confidence level.

\section{Comparison of models with data}
\label{sec:comp}

Having tuned the parameters  by fitting the near-infrared TF relation,
we now compare the resulting models to other independent observations,
to test, amongst   others, the  validity   of the star formation   and
feedback recipes  used.  As shown in  paper~I, within the  CDM context
one  can construct many physically different   models that all fit the
empirical TF relation.   For these models additional constraints, such
as the ones described below,  are required to descriminate between the
models.

\subsection{The data}
\label{sec:data}

Our calculations  yield  magnitudes  and surface  brightnesses  in the
$K$-band and  rotation velocities for  a collection of model galaxies. 
Ideally, we would  therefore compare our  models to data that consists
of   $K$-band photometry   combined  with  full  HI  rotation curves.  
Unfortunately,  such  data is currently  scarce,  and we therefore use
$B$-band photometry,  for which ample data  is available.   Although a
constant  stellar mass-to-light ratio  is a  reasonable assumption for
the $K$-band,   $\Upsilon_B^{*}$ is  expected to reveal   a systematic
trend with mass as  well as a significant amount  of scatter owing  to
variations  in dust content,  and the  ages  and metallicities  of the
stellar populations.  We convert the $K$-band luminosities and surface
brightnesses of  our  model   galaxies  to the $B$-band    adopting an
empirically   determined $B-K$    color   magnitude   relation    (see
Appendix~A).  This, at least, takes account of the systematic trend of
$B-K$ with $M_K$. Unfortunately, the  scatter is relatively large, and
we   predict errors  in $M_B$  and   $\mu_{0,B}$,  owing to  the color
conversion,  of $0.37$ and $0.47$ mag,  respectively.  A more detailed
comparison  of   our   models against data    has   to await  accurate
near-infrared photometry of a large  sample of disk galaxies for which
full HI rotation curves are available.

We compiled the following data for comparison with our models:

\begin{itemize}

\item  $M_K$, $\mu_{0,K}$, $R_d$, and $V_{\rm  flat}$ for  a sample of
  spirals in the Ursa    Major cluster from  the thesis   of Verheijen
  (1997).   We  restrict ourselves to  the  twenty-two galaxies of the
  ``unperturbed        sample''  on   which    the     TF  relation of
  equation~(\ref{TFfund}) is based.   As for our  models, we transform
  $M_K$  and  $\mu_{0,K}$  to   the $B$-band, using   the  $B-K$ color
  magnitude relation presented in  Appendix~A.  Data from  this sample
  are plotted as open circles.

\item $M_B$, $\mu_{0,B}$,  $R_d$, and $V_{\rm  flat}$ for  a sample of
  thirty-two galaxies   compiled by MB98a.  These data  are plotted as
  solid circles.

\item   $M_B$,  $\mu_{0,B}$, $R_d$, and  $V_{\rm   max}$ for ten dwarf
  galaxies from  the  sample of  van  Zee \etal  (1997).  The velocity
  measure $V_{\rm max}$ is virtually identical  to the velocity at the
  last  measured point  of   the  HI  rotation   curves, and as   such
  compatible with the velocity  measure determined in our  models (see
  paper~I). These data are plotted as solid squares.

\item $M_B$,  $\mu_{0,B}$, and $M_{\rm  HI}/L_B$ for  a sample  of one
  hundred disk galaxies of type Sb or later, compiled  by McGaugh \& de
  Blok (1997).  These data are plotted as open squares.

\end{itemize}

All data have been converted to  $h_{70}=1$, using the distance scales
quoted by the various authors. Note that this  combined set of data is
by no means complete nor  homogeneous.  In compiling  the data we have
aimed for a maximum coverage in both luminosity and surface brightness
of  disk galaxies with measured HI  rotation curves.  Therefore, it is
important to  keep in  mind that upon   comparing models to  data, one
should only focus on global properties, such as the areas of parameter
space   occupied by the  data and   the models.   Our  models are  not
constructed  to correctly predict the   local density distribution  of
data points in parameter space (i.e., the masses of our model galaxies
are sampled uniformly rather than from  a Press-Schechter formalism or
a  luminosity function). In addition,  the incompleteness and possible
observational biases in  the  data samples, combined with  the  errors
associated with   the conversion of  $K$-band models  to the $B$-band,
prevents us from testing our models in too much detail.

\subsection{Gas mass fractions}
\label{sec:gasfrac}

One of the empirical characteristics of disk  galaxies is the increase
of  their  gas mass fractions with   decreasing luminosity and surface
brightness (i.e., item~4 in  \S\ref{sec:intro}).  This is shown in the
upper  left  panels of Figures~\ref{fig:gasmag}   and~\ref{fig:gassb},
where we plot  the logarithm of   the HI mass-to-light ratio,  $M_{\rm
  HI}/L_B$, as  function  of $M_B$ and $\mu_{0,B}$,  respectively. The
remaining panels plot the  results for each of the  three models.  The
solid lines  in Figures~\ref{fig:gasmag}  and~\ref{fig:gassb}  have no
physical meaning, but  are plotted to facilitate  a  comparison of the
models with the data.

\placefigure{fig:gasmag}

The  increase of $M_{HI}/L_B$   with decreasing magnitude  and surface
brightness is remarkably well reproduced by each  of the three models.
Note  also that the  range spanned  by the  model galaxies  is in good
agreement with the data.  The HI mass-to-light ratios derived from the
models scale      with $h_{70}$, whereas    the   measured  ratios are
independent  of distance (the  HI  mass is  computed directly from the
observed flux).  The main parameter   in the models that sets  $M_{\rm
HI}/L_B$ is the  star formation parameter $Q$.   It is reassuring that
for $H_0 =  70 \kmsmpc$ and for the  value of $Q$ that was empirically
determined by Kennicutt (1989), the   models yield gas mass  fractions
that are in good agreement with observations.

\placefigure{fig:gassb}

\subsection{Dynamical mass-to-light ratios.}
\label{sec:ups}

Dynamical  mass-to-light ratios are   seen to vary systematically with
mass and surface brightness. Dwarfs and LSB galaxies have consistently
higher   mass-to-light  ratios   than   bright HSB  galaxies.    These
mass-to-light  ratios are higher than can  be explained by differences
in  stellar populations  alone,  and   indicate an increasing   ``mass
discrepancy'' with decreasing luminosity and surface brightness (i.e.,
item~3 in \S\ref{sec:intro}).

In this paper we define a characteristic mass-to-light ratio\footnote{
MB98a  defined $\Upsilon_0$ as  the global  mass-to-light ratio inside
four disk scale lengths. If we assume that $V_c(4\, R_d) \simeq V_{\rm
flat}$  and  that  disks are   pure   exponentials, our definition  of
$\Upsilon_0$ is  identical  to  that of   MB98a, apart from   a factor
$0.23$.}
\begin{equation}
\label{massdiscr}
\Upsilon_0 = {R_d \, V_{\rm flat}^2 / G  \over L_B}.
\end{equation}

\placefigure{fig:upsmag}

In Figure~\ref{fig:upsmag} we plot ${\rm log}(\Upsilon_0)$ as function
of $M_B$ for the data and our three models.   The thin line is plotted
to facilitate  a comparison, and  is   chosen to  roughly outline  the
region  that  reveals a  clear  deficit of   galaxies; low  luminosity
galaxies  always   have high  mass-discrepancies,  whereas bright disk
galaxies can  have values of $\Upsilon_0$  that span over an  order of
magnitude.  This absence of faint galaxies with low mass discrepancies
is well reproduced by  model L5, and owes largely  to the effect of SN
feedback.  The efficiency  with which SNe can blow  gas out of a  dark
halo  is larger in  less massive  galaxies.  This causes  most of  the
baryonic mass to be expelled  from low-mass halos, resulting in higher
values of $\Upsilon_0$ for  fainter galaxies.  A  possible shortcoming
of   model L5  seems  to be  the  fact  that it  predicts virtually no
galaxies  with  $\Upsilon_0 \lta 0.8$, whereas   four  of the
sixty-four galaxies    in  the  combined data    set   have values  of
$\Upsilon_0$ below this  value.    However, given the scatter  in  the
$B-K$ color magnitude  relation,  the   inhomogeneous nature  of   our
comparison sample, and  the  typical errors associated with  the data,
this apparent deficit is not significant.

In the MOND model without feedback  (model~M1), there is no reason for
fainter galaxies to have systematically higher values of $\Upsilon_0$.
Indeed  model~M1 does   not reproduce  the  observed absence  of dwarf
galaxies with   low mass discrepancies,  and   it does a  poor  job in
reproducing  the data. We  therefore included  feedback, and tuned the
parameters $\varepsilon_{\rm  SN}^0$ and $\nu$  such that the observed
`region of avoidance'   is  reproduced, yielding model~M2.   Note that
model M2 predicts, similarly to model~L5, an absence of model galaxies
with $\Upsilon_0 \lta 0.8$.

\placefigure{fig:sbvel}

A similar result is shown in Figure~\ref{fig:sbvel}, where we plot the
$B$-band central surface  brightness versus $V_{\rm flat}$.   The data
reveal   a clear  absence   of  HSB disk  galaxies  with  low rotation
velocities. In other  words, disk galaxies with  $V_{\rm flat} \lta 50
\kms$  always   have low  surface  brightness.   Again,   our DM model
reproduces this behavior remarkably  well.  Model~M1 predicts no  such
deficit, and is in  clear conflict with  the data.  Model M2, however,
in which the  two SN parameters have been  tuned to yield the observed
absence of dwarf galaxies with low mass discrepancies, is in excellent
agreement with the data.

Note  that  the observed regions   of avoidance in   both  the $M_B$ -
$\Upsilon_0$ and the $V_{\rm  flat}$ - $\mu_{0,B}$  planes are not due
to  observational biases: There is  no reason why  dwarf galaxies with
high central surface brightnesses should  be missed, whereas their LSB
counterparts are not (except for a possible problem due to star-galaxy
separation for  compact  systems at  large  distances). This therefore
seems  to be a  strong indication that  SN feedback plays an important
role   in the formation of  (disk)  galaxies, at  least   for the less
massive ones.

\placefigure{fig:upssb}

MB98a have shown that $\Upsilon_0$ is strongly correlated with central
surface   brightness.  This is  confirmed from  our  combined data set
(which includes the data of MB98a) plotted  in the upper left panel of
Figure~\ref{fig:upssb}.      The thin    solid   line corresponds   to
$\Upsilon_0^2 \propto 1/\Sigma_0^{*}$, and  is plotted with  arbitrary
normalization  for comparison.  Here    $\Sigma_0^{*}$ is the  central
surface   brightness of the    luminous,  stellar disk  in $\Lsun/{\rm
  pc}^2$.   The  data clearly     indicates  that   $\Upsilon_0^2   \,
\Sigma_0^{*} \sim {\rm constant}$ with only little scatter.

MB98a   and     MB98b  argue   that    this   $\Upsilon_0$--$\Sigma_0$
``conspiracy'' follows  naturally  from MOND,  whereas   it requires a
problematic fine-tuning in  the case of DM.  As  shown  by Zwaan \etal
(1995), $\Upsilon_0^2 \, \Sigma_0^{*} = {\rm constant}$ is required to
obtain a TF relation in  which HSB and   LSB galaxies follow the  same
relation with a slope of $b=-10$.   MB98a discuss a number of physical
processes  that  may lead to   the  observed $\Upsilon_0$ --$\Sigma_0$
``conspiracy'' under the  hypothesis of DM,  none of which they render
feasible.  The  upper right panel  of Figure~\ref{fig:upssb}, however,
shows that our  DM model reproduces the  observed ``conspiracy''  to a
good degree  of  accuracy.  Given  that the   same model  yields a  TF
relation with a slope close to $-10$  and with little scatter, this is
not too  surprising, since the $\Upsilon_0$--$\Sigma_0$ conspiracy and
the TF relation are closely connected  (see discussions in Zwaan \etal
1995 and MB98a).  We  thus disagree with  MB98a, and conclude that the
$\Upsilon_0$--$\Sigma_0$ relation is in fact fairly easily reproduced.

The lower two panels of Figure~\ref{fig:upssb}  show that the two MOND
models are also in  good agreement with the  data, something  which is
not too  surprising  given   that pure   MOND dynamics   predict  that
$\Upsilon_0^2 \propto 1/\Sigma_0^{*}$ (MB98b).

\subsection{Characteristic accelerations}
\label{sec:acc}

\placefigure{fig:xi}

Milgrom (1983b) defined the characteristic acceleration parameter
\begin{equation}
\label{xi}
\xi \equiv {V_{\rm flat}^2 \over a_0 \, R_d},
\end{equation}
which is proportional to  the ratio of the characteristic acceleration
$V_{\rm   flat}^2/R_d$  of    a  disk  and  the  characteristic   MOND
acceleration  $a_0$.  As shown by  MB98b, for an exponential disk MOND
predicts  a tight correlation between  $\xi$ and surface brightness of
the  form  $\xi \propto \Sigma_0^{1/2}$.   The   upper  left panel  of
Figure~\ref{fig:xi} plots ${\rm  log}(\xi)$ as function of $\mu_{0,B}$
for the   data.  The  thin  solid line   corresponds  to $\xi  \propto
\Sigma_0^{1/2}$,   and is plotted  with   an arbitrary normalization.  
Clearly, the  data reveals  a   narrow correlation between   $\xi$ and
central  surface brightness, with a  slope  close to that predicted by
MOND.  MB98b considered this yet another success of modified Newtonian
dynamics.

However, as we show in Appendix~B, the dark matter hypothesis predicts
the   same relation, i.e.,   $\xi   \propto \Sigma_0^{1/2}$,  and  the
observed relation between $\xi$ and  $\mu_{0,B}$ thus does not provide
a  useful test to  discriminate  between the MOND  and DM  hypotheses. 
This is also evident  from Figure~\ref{fig:xi}, which shows that  each
of  our three models yields  a relation between  $\xi$ and $\mu_{0,B}$
that is in excellent agreement with the observations. It is important,
though, to note  that under the  DM hypothesis, the  amount of scatter
can be significantly  larger than observed.   It is because  of the SN
feedback, the efficiency of which decreases with increasing halo mass,
that the amount of  scatter  in model~L5 is  as  small as it is   (see
Appendix~B for a detailed discussion).

\placefigure{fig:acc}

Additional support for MOND was recently  presented by McGaugh (1998),
who showed  that  current observations  hint  to  the  presence  of  a
characteristic  acceleration, the  main assumption  on  which MOND  is
based.   The upper panels   of Figure~\ref{fig:acc} plot data,  kindly
provided by   Stacy   McGaugh, of   the cumulative   mass  discrepancy
$\Upsilon$  against radius $R$  (panels on the right), against orbital
frequency   $\omega = V_{\rm  rot} /R$   (middle panels), and  against
centripetal  acceleration $\alpha \equiv  V_{\rm  rot}^2/R$ (panels on
the right). Here $\Upsilon$ is defined as
\begin{equation}
\label{massdiscrep}
\Upsilon(R) = {R \, V_{\rm  rot}^2(R) \over G   \, [M_{*}(R) +  M_{\rm
    cold}(R)]},
\end{equation}
and  corresponds to the  ratio of  total  mass (based on the Newtonian
equations) to luminous mass (stars  plus cold gas)  within radius $R$. 
Each data point represents  one  resolved measurement in the  rotation
curve of a disk galaxy.  The galaxies in  the sample span a wide range
in  luminosities  and surface brightnesses    and were taken  from the
compilation of Sanders (1996) and de Blok \& McGaugh (1998).

The plot of  $\Upsilon$ versus $R$  reveals that the  mass discrepancy
sets in at a wide range of different radii; individual rotation curves
are easily discerned. This scatter is reduced when plotting $\Upsilon$
versus the circular frequency, and becomes minimal when plotted versus
the acceleration $\alpha$.  The data  clearly reveal that in each disk
galaxy,  the mass discrepancy starts  to  become significant below  an
acceleration of $\sim 10^{-10} \mss$, as predicted by MOND.

The lower two rows of   panels of Figure~\ref{fig:acc} show the   same
plots for the  two  MOND models.  Here   we have randomly  selected 40
galaxies from models~M1 and~M2, and for each model galaxy, we computed
$\Upsilon$ at 15 radii, sampled uniformly between $R=0$ and the radius
at which  the projected  HI column density  is equal  to $10^{20} {\rm
  cm}^{-2}$.   The two MOND models reproduce  the data extremely well,
which is  to be  expected, given  the  presence of the  characteristic
acceleration $a_0$ and the fact that MOND was specifically designed to
fit rotation curves. More     remarkably,  model L5 also    yields   a
characteristic  acceleration  (see     panels   in   second   row   of
Figure~\ref{fig:acc}).  Although the  agreement  with the data is  not
quite  as  good as   in  the  case  of   models~M1 and~M2,   the  main
characteristic of the data, namely  the  minimization of scatter  when
$\Upsilon$   is plotted versus   $\alpha$,  is reproduced  well.  Most
importantly, this  is achieved without  any fine-tuning of  our model;
after setting the   feedback   parameters to  fit the   slope   of the
empirical  TF  relation,  the same   model  reveals a   characteristic
acceleration.  This is in clear contradiction with McGaugh (1998), who
argued that this phenomenology can  not be reproduced  with DM models. 
Not only is this a remarkable success  of our DM  model, it is also an
argument against an argument in favor of MOND.

Model L5  predicts mass discrepancies which are  slightly too large in
the range $3 \times  10^9 \lta \alpha  \lta 10^{10} \mss$.  The origin
of this small deficit is discussed in the following \S.

\subsection{Understanding the scaling relations}
\label{sec:understand}

Why can our DM  models easily account for the $\Upsilon_0$--$\Sigma_0$
``conspiracy'', whereas MB98a have  argued  that no feasible  solution
exists?  Combining equations~[\ref{massdiscr}]  and~[\ref{xi}],    and
taking into account that the  total luminosity of an exponential  disk
scales as $L \propto \Sigma_0 R_d^2$), one obtains $\Upsilon_0 \propto
\xi \,  \Sigma_0^{-1}$.  Since, as shown  in  Appendix~B, we have $\xi
\propto   \Sigma_0^{1/2}$  one   thus   automatically  predicts   that
$\Upsilon_0   \propto \Sigma_0^{-1/2}$;  i.e.,     {\it  there  is  no
  conspiracy}.   However, the  fact   that   the scatter  around   the
$\Upsilon_0$--$\Sigma_0$ relation is so  small is not trivial, as  the
scatter in the $\xi$--$\Sigma_0$ relation  can be fairly large.  It is
only  because of  our  particular feedback model,   which we  tuned to
reproduce   the empirical TF  relation,   that   the  scatter in   the
$\xi$--$\Sigma_0$      relation,    and  therewith     in          the
$\Upsilon_0$--$\Sigma_0$ relation,  is so small  (see Appendix~B).  We
consider this another major succes for our feedback model.

From the scaling relations listed above  one derives that $\xi \propto
\Upsilon_0^{-1}$.  We thus  expect a  narrow   correlation between the
characteristic    accelerations   and  mass-to-light  ratios  of  disk
galaxies.   Therefore, the  fact that our    DM model is  succesful in
reproducing the  narrow correlation between  {\it local} accelerations
and mass-to-light ratios is not too surprising.

\section{Rotation curve shapes}
\label{sec:rcs}

The   rotation curves (hereafter RCs) of   disk galaxies  are the most
direct  indicators of a discrepancy  between the luminous mass and the
inferred dynamical mass  for  disk galaxies.   The shapes of  rotation
curves therefore provide  a clean laboratory  for comparing models  of
gravitational  dynamics.  Historically, astronomers have explained the
detailed    RCs of  spiral   galaxies  by  invoking  the presence   of
non-luminous matter  (see reviews in  Sancisi \&  van Albada 1987, and
Ashman  1992), while preserving  the $r^{-2}$  force law of  Newtonian
gravity.  More recently, some astronomers  have  used MOND to  explain
the observed kinematics without requiring the  presence of dark matter
(e.g.\ Kent 1987; Lake 1989; Begeman \etal 1991; Sanders 1996; de Blok
\&  McGaugh 1998; Sanders \&   Verheijen 1998).  Both approaches  have
found  nearly  equal success over  a wide   range  of luminosities and
surface   brightnesses,    well  beyond   the  luminous,  high-surface
brightness  galaxies for   which the  theories  of  DM  and  MOND were
developed.

Recently, an important problem for the CDM picture has emerged from HI
studies of dwarf galaxies.  The slowly rising rotation curves observed
in these systems  suggest a (close to) constant  density core in their
dark matter   halos,   inconsistent with the  steeply  cusped  density
profiles   predicted  for CDM (Flores  \&   Primack  1994; Moore 1994;
Burkert 1995; Burkert \& Silk 1997; Stil 1999,  but see Kravtsov \etal
1998).  Whether MOND is consistent with the RCs  of dwarf galaxies, is
an issue that is  currently  still under  debate (Lake  1989;  Milgrom
1991; S\'anchez-Salcedo \& Hidalgo-G\'amez 1999).

Recent studies  have shown that more  massive LSB disk galaxies reveal
rotation curves that are similar to those  of dwarf galaxies (e.g., de
Blok, McGaugh \& van der  Hulst 1996;  van  Zee \etal 1997;  Pickering
\etal 1997), suggesting that the problem with  the density profiles of
dark matter halos  is not limited  to low-mass systems.  Indeed, MB98a
have  argued that the  observed  rotation curve   of LSB galaxies  are
inconsistent  with the NFW  profile. MOND, however,  has been shown to
yield  remarkably good fits  to the  RCs of  LSB  galaxies (de Blok \&
McGaugh 1998;  Sanders \& Verheijen  1998).  It is important  to note,
however, that the majority of the data on the  RCs of the more massive
LSB galaxies is severely affected by beam  smearing (see e.g., Swaters
1999).  Van den Bosch, Robertson  \& Dalcanton (1999) have shown  that
once these effects are taken into account, the RCs of the more massive
LSB galaxies  are in excellent agreement  with dark  matter halos with
steep central cusps.

\subsection{A detailed comparison}
\label{sec:comparison}

\placefigure{fig:rc}

Given the unique opportunity to discriminate between DM and MOND based
on the RC shapes of disk galaxies, we now compare the RCs predicted by
our DM and MOND models in some detail.  Using our models which best fit
the $K$-band TF relation (L5 and M1), we  have selected model galaxies
that  are  photometrically similar (same  scale  length, same absolute
magnitude, and same  central surface  brightness), and computed  their
RCs.  In Figure~\ref{fig:rc}  we  compare four sets of  model galaxies
that span  four  magnitudes in  both surface  brightness and mass (see
Table~\ref{tab:rcgal} for   the  parameters   of each  of    the eight
galaxies).

For the high surface brightness galaxies, (d1, m1, d3, and m3) the RCs
for the MOND and the DM models are very similar. For the LSB galaxies,
however,   the  galaxies  from  the DM  model    predict  RCs that are
significantly steeper in the inner parts.  This is the reason why dark
halo fits to several LSB galaxies have been  claimed to fail, and owes
to the steep central cusp ($r^{-1}$ for the NFW profile used in our DM
models) of   the dark halo.  This also   explains why $\Upsilon(R)$ is
slightly overestimated for accelerations at around $3 \times 10^9 \lta
\alpha \lta 10^{10} \mss$  (see  Figure~\ref{fig:acc}) as compared  to
the    observations.   Note,  however,  that    the   data  on   which
Figure~\ref{fig:acc} is based suffers  strongly from the beam-smearing
effects eluded to  above,  and it remains  to  be seen  whether it  is
actually inconsistent with the predictions  from our DM model (see van
den Bosch \etal 1999).

The steepness of the central part of the RCs can  be quantified by the
parameter $R_{3/4}$,   which  is defined as the   radius  at which the
circular velocity equals 75 percent of $V_{\rm flat}$.  MB98a obtained
$R_{3/4}$ for a number of disk  galaxies with reasonably well resolved
HI rotation curves, and found $R_{3/4}/R_d$  to strongly increase with
decreasing luminosity  and surface brightness.  They compared  this to
predictions  based on the models of   Dalcanton \etal (1997; hereafter
DSS97), which  are similar to  the models presented here. According to
MB98a, the   models of DSS97  predict  an opposite behavior  from that
observed,   with   $R_{3/4}/R_d$  {\it decreasing}    with  decreasing
magnitude, and they  use  this as strong  evidence against  DM models.

\placefigure{fig:rcrad}

To further test this, we compute  $R_{3/4}/R_d$ for the model galaxies
in each of our three models. Results are plotted  as function of $M_B$
and $\mu_{0,B}$ in Figure~\ref{fig:rcrad}. The first characteristic to
notice is that  none of the models  reveals a significant  increase of
$R_{3/4}/R_d$ with decreasing magnitude.    The MOND models   reveal a
modest increase of $R_{3/4}/R_d$  with decreasing surface  brightness,
but they reach a maximum at $R_{3/4} \simeq 1.25 R_d$. The galaxies of
model L5 never reach values of $R_{3/4}/R_d$ in  excess of $\sim 0.8$. 
Both models are therewith in clear contradiction with the observed RCs
presented  by MB98a, which reach  values  as high as $R_{3/4}/R_d \sim
3$.  The most likely explanation for  this discrepancy is, once again,
beam smearing, which tends to   over-predict $R_{3/4}/R_d$ (see  e.g.,
Blais-Ouellette,  Carignan  \& Amram   1998). Data of   higher spatial
resolution is required to test this further. However, given the modest
difference between the  predictions from  MOND  and the DM models,  it
seems unlikely that $R_{3/4}/R_d$  will allow to discriminate  between
the two scenarios

Not only are the  results presented here  inconsistent with data, they
also disagree with the  curves  plotted by MB98a  which they  claim to
represent the   model  predictions  from DSS97.  However,  we  do  not
understand  how MB98a have  obtained  these predictions.   Part of the
answer may lie in the fact that MB98a quote that they used models with
$L \propto \Sigma_0^{1/3}$  ``as  predicted by DSS97''.   However, the
models of DSS97   predict $\Sigma_0  \propto  L^{1/3}$.  Nevertheless,
even with  this  erroneous interpretation of   DSS97's  models, we are
unable to  reproduce the predictions  of MB98a.  The results presented
here suggest that neither MOND nor  the DM hypothesis predict a strong
dependence   of  $R_{3/4}/R_d$    on either    luminosity  or  surface
brightness. Together with  the fact that the MOND  and DM models yield
results that differ  only modestly,  we conclude  that they both   are
equally  (in)consistent   with the  data.    The  observed  values  of
$R_{3/4}/R_d$ can not be used to simply rule against the DM hypothesis
only.

\section{Conclusions}
\label{sec:concl}

We have  presented detailed models  for the formation of disk galaxies
in both a CDM  and a MOND universe.  In  the case of DM, the structure
of the disk   is governed by the  mass   and angular momentum  of  the
proto-galaxy.  For MOND, however, the  distribution of angular momenta
of proto-galaxies is not  known,  and we  have instead  assigned scale
lengths  to the disks  that  are in agreement  with observations.   In
addition to these recipes that determine the structure and dynamics of
the disks,  we include recipes that describe  how (part of) the gas is
transformed into stars over    the lifetime of  the galaxy.    We take
account of a stability  related threshold  density for star  formation
and feedback  from supernovae.  The models  have been tuned to fit the
observed $K$-band TF  relation, and compared to numerous observations.
Both models   do  remarkably well  in  reproducing a  wide variety  of
observations of disk galaxies that span several orders of magnitude in
both luminosity and surface brightness.

To more easily compare  the different models  a summary of the results
is given in  Table~\ref{tab:results},  where we indicate, for  each of
the observational constraints discussed  in  this paper, whether   the
models are consistent with the data or  not. In the  case of DM (model
L5), SN feedback is required to yield a TF slope as steep as observed.
Although the  MOND  model  without   feedback (M1) is   in   excellent
agreement  with the empirical TF  relation, it does  not reproduce the
observed    deficit   of   HSB     dwarf    galaxies     with    small
mass-discrepancies. This  can be remedied  by introducing  SN feedback
(model M2), but at the  cost of a TF relation  which is too steep  and
reveals  an amount of scatter that  is only marginally consistent with
the data.  This is  a serious problem  for  MOND; since  pure dynamics
already  predicts  a  TF relation  as   steep  as observed,  it leaves
virtually no  room  for   any other  galaxy characteristics   to  vary
systematically with mass.

The DM model is consistent  with all observational constraints against
which we have tested it.  Consequently,  we strongly disagree with the
picture that emerges from the  literature (for example, MB98a,  MB98b,
and McGaugh  1998),   that the DM  hypothesis   suffers  from numerous
serious fine-tuning problems,  which do not seem  to have a  clear-cut
solution,   whereas MOND is free  from  such problems   and capable of
fitting virtually everything.  We  have shown here   that once the  DM
model is tuned to fit  the slope of the  TF relation, it automatically
passes all  the  tests devised by  MB98a  and McGaugh (1998)  to argue
against it. In particular, our DM model reproduces

\begin{itemize}

\item A close correlation    between global mass-to-light ratio   and
  surface brightness, such that HSB  and LSB galaxies follow the  same
  TF relation without a systematic offset.

\item The presence of a characteristic  acceleration as observed. 

\item A close relation between the characteristic acceleration, $\xi$,
  and central   surface    brightness  of  the  form    $\xi   \propto
  \Sigma_0^{-0.2}$. 

\end{itemize}

This is a remarkable  result: there is  no obvious  reason why the  DM
model would  reveal a characteristic acceleration,  unlike in the case
of  MOND,  where it is integral   to the  theory.   Furthermore, it is
encouraging that  the same feedback  parameters that yield the correct
TF slope, result in an amount  of scatter around the $\xi$--$\Sigma_0$
relation that is in    excellent agreement with observations,   and in
addition  explains  the observed  absence of   HSB  dwarfs with  small
mass-discrepancies.

Of  the list of observational facts   about disk galaxies presented in
\S\ref{sec:intro}, the upper limit  to the observed surface brightness
of  disk  galaxies  (item~5) has   not been  addressed by our  models.
Several studies  have  shown,  however, that  in  the  DM picture  the
presence of a  maximum central surface  brightness of disk galaxies is
related to stability  arguments (DSS97; Mo \etal  1998; Scorza \&  van
den  Bosch 1998).  In \S\ref{sec:tfmond}  we have shown  that the same
stability  argument also yields an  upper limit on the central surface
brightness of  disks under the hypothesis of  MOND (cf. Milgrom 1989).
Henceforth,  both the DM and  the MOND models  are consistent with the
observational constraint of item~5, as long as disk stability is taken
into account.

The main problem for CDM is related to rotation curve shapes.  We have
compared predicted rotation  curve shapes within  the context of  both
MOND and CDM.  Only in  the case of  LSB systems do the two  scenarios
yield RCs that    are significantly different.  Several studies   have
pointed  out that dark halos with  a  steep cusp are inconsistent with
the central rotation  curves of LSB and  dwarf galaxies.  This problem
is also evident from the fact that at  accelerations of $\sim 10^{-10}
\, h_{70} \, \mss$ the DM model predicts mass-to-light ratios that are
slightly too high.  However, it is important to  realize that most data
on the more  massive LSB disks is  severely affected by beam-smearing. 
This tends to     underestimate  the central gradients   of   rotation
velocities, especially  in galaxies that have a  central hole in their
HI distribution.   When  beam smearing  is taken into  account, the HI
rotation curves of massive LSB galaxies are consistent with dark halos
that follow a NFW density profile (van den  Bosch \etal 1999). Further
indications that beam  smearing plays an  important role  comes from a
comparison of the  ratio $R_{3/4}/R_d$.    Both  the DM and  the  MOND
models  predict ratios  that  are lower than  observed,  which is most
likely due  to beam smearing.   Therefore,   in order to  discriminate
between MOND  and DM, high   resolution  rotation curves of  LSB  disk
galaxies  are  required.  Such data is    currently only available for
dwarf galaxies, which, because  of their relative proximity, have been
observed with high    spatial  resolution. It has  been   demonstrated
convincingly  that the rotation  curves  of these low-mass systems are
inconsistent with centrally cusped dark matter halos.  It is currently
still under debate  whether MOND can  fit these rotation curves. If it
can, this is where MOND has a  clear advantage over CDM, unless (close
to) constant density cores can be produced in  dark halos that form in
a CDM  Universe (see e.g., Navarro,  Eke \& Frenk 1996, Kravtsov \etal
1998, and Bullock \etal 1999 for possible solutions).

One  might argue  that within  the DM  scenario one can  fit basically
anything as long as there are a sufficient  number of free parameters.
In that  respect it is  important to realize  that model L5 has only a
very limited number of truly free parameters.  Where possible, we have
used parameters that  have  either empirically determined  values,  or
that are otherwise  constrained:  the dark halo properties   are taken
from high   resolution   numerical  simulations   combined with    the
Press-Schechter  formalism, $\Upsilon_K^{*}$ is constrained by stellar
population  models, the star formation  recipe uses values for $Q$ and
the  Schmidt  law   that    have been determined   empirically,    and
$\alpha_{\rm  crit}$ is  constrained  by numerical  simulations.  This
leaves only   $\varepsilon_{\rm  SN}^0$   and  $\nu$  as   real   free
parameters.  Tuning these two parameters  to obtain a TF relation with
the observed   slope  of $b=-10.5$,  the   model  predicts  gas   mass
fractions, characteristic  accelerations, an $\Upsilon_0$ - $\Sigma_0$
``conspiracy'', and global  mass-to-light    ratios which are   in
excellent agreement   with observations, without  additional tuning of
the parameters.   Furthermore,  we  note  that  while MOND   is  often
presented as being nearly free  of fine-tuning, in  order to match the
systematic properties of disk galaxies  it is necessary to adjust  the
MOND feedback parameters to the same degree as required for DM.

Probably the most amazing aspect of the models presented here, is that
the DM and MOND models are so very similar.  However, both MOND and DM
were constructed to fit the rotation curves of disk galaxies. The fact
that both theories correctly  predict many other properties, which are
themselves   closely  related to  the   internal  dynamics  (i.e.,  TF
relation,  gas  mass fractions   that are   set  by stability  related
threshold densities,  SN   feedback whose  efficiency   depends on the
escape velocity, etc), should therefore not be seen as too remarkable.
It has  often  been argued that  even  if MOND  turns   out to not  be
correct, one should provide  an  explanation  as to  why it  fits  the
properties of galaxies  so well.   The  demonstrations in  this  paper
suggest that {\it any} theory which yields stable disks and fits their
rotation curves, would  probably perform as well as  any of the models
presented here.


\acknowledgments

This work has benefited greatly from discussions  with George Lake. We
are grateful to Stacey  McGaugh for sending us  his data in electronic
format, and to  the anonymous referee for  his suggestions that helped
to improve the paper.  FvdB   was supported  by NASA through    Hubble
Fellowship grant \#   HF-01102.11-97.A awarded by the Space  Telescope
Science  Institute, which is operated by  AURA for NASA under contract
NAS 5-26555.


\clearpage

\begin{appendix}

\section{A. The $B-K$ color magnitude relation for spiral galaxies}
\label{sec:AppA}

In order to compare our models, which yield $K$-band magnitudes and
surface brightnesses, to $B$-band data, we require an estimate of
$B-K$ colors for disk galaxies. 

\placefigure{fig:colors}

To determine an empirical  $B-K$ color magnitude relation  we compiled
apparent magnitudes in $B$ and $K$ as well as distances for a total of
139 spiral  galaxies of type Sb  or later from  the following sources:
forty-two  Ursa-Major spirals   from the sample  of  Verheijen (1997),
sixty-one galaxies from the sample of de Jong \& van der Kruit (1994),
and thirty-six  galaxies    from the  sample   of  edge-on spirals  of
Dalcanton  \etal (1999).   Absolute  magnitudes were computed assuming
$H_0 =  70  \kmsmpc$  and  using  the  distances as   quoted  by these
authors. Finally,  we used the reddening  maps of Schlegel, Finkbeiner
\& Davis  (1998), with  the $R_V =   3.1$ extinction law  of  Cardelli,
Clayton \& Mathis (1989) and O'Donnell (1994), to correct for external
extinction.

The resulting  $B-K$ color magnitude is shown  in the upper left panel
of Figure~\ref{fig:colors}. The solid line is the best linear fit with
\begin{equation}
\label{colmag} 
B-K = -0.66 - 0.18 ( M_K - 5 {\rm log} h_{70}).
\end{equation}
The   upper right    panel plots the    predicted $B$-band  magnitude,
$M_B^{\rm  pred} = 0.82\,M_K^{\rm obs}   -  0.66$ versus the  observed
value (for $h_{70}  = 1$).  The data  covers over eight  magnitudes in
$M_B$  and reveals that $M_B^{\rm pred}$  is a fairly good estimate of
the observed     $B$-band magnitude (with   a  standard   deviation of
$\sigma_M = 0.37$ mag).

The lower left panel of  Figure~\ref{fig:colors} plots the $B-K$ color
of the central surface brightness of the disks, versus $\mu_{0,K}$ for
the data of Verheijen (1997)   and de Jong \&  van  der Kruit (1994).  
Unfortunately,  at   this time no   central  surface  brightnesses are
available for the data of Dalcanton \etal (1999).  Note that $B-K$ and
$\mu_{0,B} -  \mu_{0,K}$  are generally  not the  same, since the disk
scale  lengths  derived  from the  $B$  and  $K$-band data can  differ
considerably   (see e.g., de Jong  1996b;   Verheijen 1997). The lower
right panel  of Figure~\ref{fig:colors}  plots the   predicted central
surface brightness in the $B$-band, $\mu_{0,B}^{\rm pred}$, versus the
observed value.    Here  $\mu_{0,B}^{\rm  pred}$   is calculated  from
$\mu_{0,K}^{\rm    obs}$     and  the   color-magnitude    relation of
equation~(\ref{colmag}).  Note that  the agreement  is reasonable, but
that the  scatter is, with $\sigma_{\mu}  = 0.47$ mag, slightly higher
than for the absolute magnitudes.


\section{B. The characteristic acceleration parameter $\xi$}
\label{sec:AppB}

In \S\ref{sec:acc} we  show that both the  DM and MOND  models yield a
narrow correlation between the central surface  brightness of the disk
and  the  parameter  $\xi$  defined by  equation~(\ref{xi}).   Here we
investigate the reason for this narrow relation.

In the case of  MOND, $V_{\rm flat}^4 = G\,  a_0 \, M_{\rm gal}$, and,
if we assume  that  the galaxy  is  a pure   exponential disk,  it  is
straightforward to show that
\begin{equation}
\label{ximond}
{\rm log}\left( \xi_{\rm MOND} \right) = -{1 \over 5}(\mu_{0,B} -
27.052) + {1 \over 2} \, {\rm log}\left( {2 \, \pi \, G \over a_0}
  \Upsilon_{\rm gal} \right),
\end{equation}
(see MB98b), with  $\Upsilon_{\rm gal}$ the  total mass-to-light ratio
of the galaxy.

In order to obtain a similar expression under the hypothesis of DM, we
make the  simplifying assumption that  dark  halos are represented  by
isothermal spheres, and that $V_{\rm  flat} = V_{200}$.  Upon  writing
$M_d    = \epsilon_{\rm  gf}     \, f_{\rm  bar}    \, M_{200}$   (cf. 
equation~[\ref{tfdm}]), and since
\begin{equation}
\label{appmass}
M_{200} = {V_{200}^3 \over 10 \, G \, H_0},
\end{equation}
we obtain
\begin{equation}
\label{xidm}
\xi_{\rm DM} = \sqrt{20 \, \pi \, G \, H_0 \, V_{200} \, \Sigma_0
  \over a^2_0 \, \epsilon_{\rm gf} \, f_{\rm bar}}.
\end{equation}
If we now set $H_0 = 70 \kms$ and $f_{\rm bar} = 0.085$ (as for model
L5, see Paper~I), we can write
\begin{equation}
\label{xicompare}
{\rm log}\left( \xi_{\rm DM} \right) = {\rm log}\left( \xi_{\rm MOND}
\right) - 0.11 + {1 \over 2} {\rm log}\left( {V_{200} \over 250 \kms}
\right) - {1 \over 2} {\rm log}(\epsilon_{\rm gf}).
\end{equation}

Under the DM  hypothesis one thus  expects a similar  relation between
$\xi$ and $\mu_0$ as under MOND, apart from an offset which depends on
$V_{200}$ and   the galaxy formation  efficiency  $\epsilon_{\rm gf}$.
Since disks with  the same central surface  brightness can be embedded
in        halos  with     different      rotation   velocities    (cf.
Figure~\ref{fig:sbvel}), one expects a reasonable amount of scatter; a
variation in $V_{200}$ of a factor five, yields a scatter of $0.35$ in
${\rm  log}(\xi)$. However, this can be  significantly reduced when SN
feedback is important; the feedback efficiency will be larger, leading
to smaller values of $\epsilon_{\rm gf}$  in systems with lower values
of $V_{200}$, thus reducing the amount of scatter. Indeed, the scatter
revealed  by  model~L5  (see  upper  panel of Figure~\ref{fig:xi})  is
somewhat larger  than in the  case of  MOND, but significantly smaller
than $0.35$ and therewith consistent with the data.

We   thus   conclude   that   the  observed   relation    $\xi \propto
\Sigma_0^{1/2}$ does  not discriminate  between  the  DM and the  MOND
hypotheses.  However, the small   amount  of scatter observed   may be
considered  evidence for  efficient SN feedback,   at least within the
context of DM.

\end{appendix}


\ifsubmode\else
\baselineskip=10pt
\fi


\clearpage


\ifsubmode\else
\baselineskip=14pt
\fi


\newcommand{\figcaptf}{A  comparison  of  the  observed   $K$-band  TF
  relation with our  three models. The  upper  left panel plots,  with
  open   circles,  the  twenty-two    Ursa-Major  spirals    from  the
  ``unperturbed sample'' of Verheijen  (1997). The thick solid line is
  the best fitting TF relation (equation~[\ref{TFfund}]). This line is
  reproduced in each  of the panels. The upper  right  panel plots the
  results for our best-fitting DM  model (model L5), whereas the lower
  two panels correspond to the two MOND models discussed  in the text. 
  The value of  the slope,  $b$, of  the  best-fitting TF relation  is
  indicated in each panel.\label{fig:tf}}

\newcommand{\figcapgasmag}{The  HI     mass-to-light   ratios, $M_{\rm
    HI}/L_B$, as  function of  total magnitude.  The  upper left panel
  plots the data for a sample of one  hundred disk galaxies of type Sb
  or later compiled  by McGaugh \& de Blok  (1997), and clearly reveal
  an increase  of gas mass fractions  with decreasing luminosity.  The
  remaining   three panels  plot  the results  for  each  of the three
  models, as labeled. The two thin lines have no physical meaning, but
  are  plotted to facilitate a   comparison  between models and data.  
  Luminosities   and magnitudes for the     model galaxies have   been
  converted from  the $K$-band to the $B$-band   using the $B-K$ color
  magnitude relation presented  in  Appendix~A.  Given the   errorbars
  associated with  the data, and with the  color conversion, all three
  models are  consistent  with   the data, and  nicely   reproduce the
  increase of $M_{\rm HI}/L_B$ with decreasing luminosity.
  \label{fig:gasmag}}

\newcommand{\figcapgassb}{Same as Figure~\ref{fig:gasmag}, except that
  now  $M_{\rm  HI}/L_B$    is  plotted versus  the   central  surface
  brightness of the  disk, $\mu_{0,B}$. Again,  the thin lines have no
  physical  meaning but   are  plotted  to facilitate  a   comparison. 
  Clearly, the three models are all consistent  with the data and with
  each other.\label{fig:gassb}}

\newcommand{\figcapupsmag}{The characteristic   global   mass-to-light
  ratio   $\Upsilon_0$ (equation~[\ref{massdiscr}])  as function    of
  absolute $B$-band magnitude.  The  data (upper left panel) are taken
  from   Verheijen (1997; open circles),    McGaugh \& de Blok (1998a;
  solid circles), and van Zee \etal (1997; solid squares).  Details on
  the data can be found in  \S\ref{sec:data}.  The data clearly reveal
  an absence of faint galaxies with  low values of $\Upsilon_0$, i.e.,
  low luminosity systems always  have a large  mass discrepancy.  This
  is reproduced nicely  by model~L5 (upper right  panel),  which is in
  reasonable agreement with the data (see the text for a more detailed
  discussion). Model~M1, the MOND  model without SN feedback, however,
  is  clearly inconsistent with   the data,  in  that it   predicts no
  deficit  of low luminosity systems  with low values of $\Upsilon_0$. 
  In   model~M2, we   have  included  SN    feedback,  and  tuned  the
  corresponding parameters to reproduce the data.  The thin solid line
  is plotted to  facilitate  a comparison,  and  is chosen to  roughly
  outline the boundary of the data.\label{fig:upsmag}}

\newcommand{\figcapsbvel}{The central  surface brightness of the disks
  ($\mu_{0,B}$) as function of the  rotation velocity at the flat part
  of the rotation curve, $V_{\rm flat}$.  The symbols for the data are
  the same as  in Figure~\ref{fig:upsmag}.   The  data reveal  a clear
  absence of HSB   galaxies with  low  rotation velocities.   This  is
  remarkably   well reproduced  by models~L5 and   M2  (both of  which
  include SN feedback), whereas model~M1 (no SN  feedback) is in clear
  contradiction with  the data.   The  thin solid   line is chosen  to
  roughly outline   the boundary  of  the data,   and has   no further
  physical meaning. Note that the absence of  observed galaxies in the
  upper left part of the diagram is not  related to observational bias
  effects. Rather, it seems to indicate that SN feedback has played an
  important role in shaping low mass disk galaxies.
\label{fig:sbvel}}

\newcommand{\figcapupssb}{Same as Figure~\ref{fig:upsmag}, except that
  we now plot the characteristic  mass-to-light ratio $\Upsilon_0$  as
  function  of   the  disk's   $B$-band   central   surface brightness
  $\mu_{0,B}$.   The  data reveal   a  narrow  correlation,  which  is
  extremely  well reproduced by  each  of the three  models.  The thin
  solid  line corresponds  to  $\Upsilon_0^2  \,  \Sigma_0^{*} =  {\rm
    constant}$, and is plotted with  arbitrary normalization. See text
  for a detailed discussion.
\label{fig:upssb}}

\newcommand{\figcapxi}{The characteristic acceleration $\xi$      (see
  equation~[\ref{xi}])  as   function of central   surface  brightness
  $\mu_{0,B}$.  The data and each of  the three models reveal a narrow
  correlation, consistent with $\xi  \propto \Sigma_0^{1/2}$ (the thin
  solid lines, plotted with arbitrary  normalization).  The origin  of
  this relation is discussed in Appendix~B.\label{fig:xi}}

\newcommand{\figcapacc}{The   ratio    of  total  to   luminous  mass,
  $\Upsilon(R)$  (equation~[\ref{massdiscrep}]), as function of radius
  $R$ (panels on   the   left), orbital frequency   $\omega$   (middle
  panels),  and  centripetal acceleration   $\alpha$  (panels  on  the
  right).  The upper  panels plot  the  data for thirty galaxies  from
  McGaugh (1998).  These clearly reveal  that the amount of scatter in
  $\Upsilon$  is minimized  when  plotted versus  $\alpha$,  therewith
  indicating the   presence  of  a characteristic   acceleration. This
  character of  the data is  extremely well  reproduced by the models.
  Whereas this  is not surprising for the  MOND models (M1 and M2), it
  is a remarkable success for the DM model.\label{fig:acc}}

\newcommand{\figcaprc}{Rotation curves for  model galaxies plotted up
  to the radius at  which the HI column  density reaches $10^{20} {\rm
  cm}^{-2}$.  Thick solid lines correspond to  the galaxies from model
  M1 (MOND), whereas  the thick dashed lines  are the  rotation curves
  for   galaxies from  model  L5  (CDM).  The thin   dotted lines  and
  long-dashed lines correspond to the contributions  from the disk and
  the dark halo,   respectively.   Each panel compares  the   rotation
  curves of two galaxies   that have virtually   identical photometric
  properties  (see Table~2),  and which  are indicated  in each panel.
  The four sets  of galaxies span  four magnitudes  in both luminosity
  and surface  brightness. Note  that  the difference between  the  DM
  rotation curves and those for MOND  increase with decreasing surface
  brightness.\label{fig:rc}}

\newcommand{\figcaprcrad}{The   ratio of  $R_{3/4}$,  defined  as  the
  radius where the RC reaches a velocity 75 percent of $V_{\rm flat}$,
  to   the disk scale  length  $R_d$ as  function of absolute $B$-band
  magnitude  (upper panels)  and  central $B$-band surface  brightness
  (lower panels) for each of the three models. Rotation curves in MOND
  are shallower  (i.e., larger $R_{3/4}/R_d$)  than in the case of DM,
  but only  marginally so.  Note that there  is  only a  very marginal
  increase of $R_{3/4}/R_d$   with decreasing luminosity and   surface
  brightness, in clear contradiction with the data presented in MB98a.
  In addition, both the DM  and the MOND  models, reveal a clear upper
  limit   of $R_{3/4} \lta   0.8 R_d$  and   $R_{3/4}  \lta 1.25  R_d$
  respectively.  This again  is in clear contrast  to the data,  which
  shows galaxies  with $R_{3/4}/R_d$ as high  as $3$.  The most likely
  explanation for this inconsistency seems that the data have not been
  properly    corrected   for   beam-smearing    (see  discussion   in
  text).\label{fig:rcrad}}

\newcommand{\figcapcolors}{The upper   left panel  plots  $B-K$ colors
  versus absolute $K$-band magnitude for an ensemble of disk galaxies.
  Data is taken  from Verheijen (1997; solid circles),  de Jong \& van
  der Kruit (1994; open circles), and Dalcanton \etal (1999; crosses).
  All galaxies have been  converted to  a  common distance  scale with
  $H_0 = 70  \kmsmpc$ and have been  corrected for external extinction
  in a  consistent  way, using the  reddening  maps of  Schlegel \etal
  (1998).  No  correction for internal extinction  has been applied to
  the  data.  The thick  solid line  in the upper   left panel is  the
  best-fit  linear color magnitude relation (equation~[\ref{colmag}]).
  The upper right panel plots the predicted $B$-band magnitudes (based
  on the observed $K$-band magnitude  and the best fitting $B-K$ color
  magnitude  relation) versus the  observed  values.  The agreement is
  reasonable with a standard deviation around the line $M_B^{\rm pred}
  = M_B^{\rm  obs}$ (thin line) of  $0.37$ mag.  The lower  two panels
  plot  similar  figures, but for   the  central surface brightnesses,
  rather than the absolute    magnitudes (see Appendix~A  for  further
  details).\label{fig:colors}}


\ifsubmode
\figcaption{\figcaptf}
\figcaption{\figcapgasmag}
\figcaption{\figcapgassb}
\figcaption{\figcapupsmag}
\figcaption{\figcapsbvel}
\figcaption{\figcapupssb}
\figcaption{\figcapxi}
\figcaption{\figcapacc}
\figcaption{\figcaprc}
\figcaption{\figcaprcrad}
\figcaption{\figcapcolors}
\clearpage
\else\printfigtrue\fi

\ifprintfig

\clearpage
\begin{figure}
\epsfxsize=14.0truecm
\centerline{\epsfbox{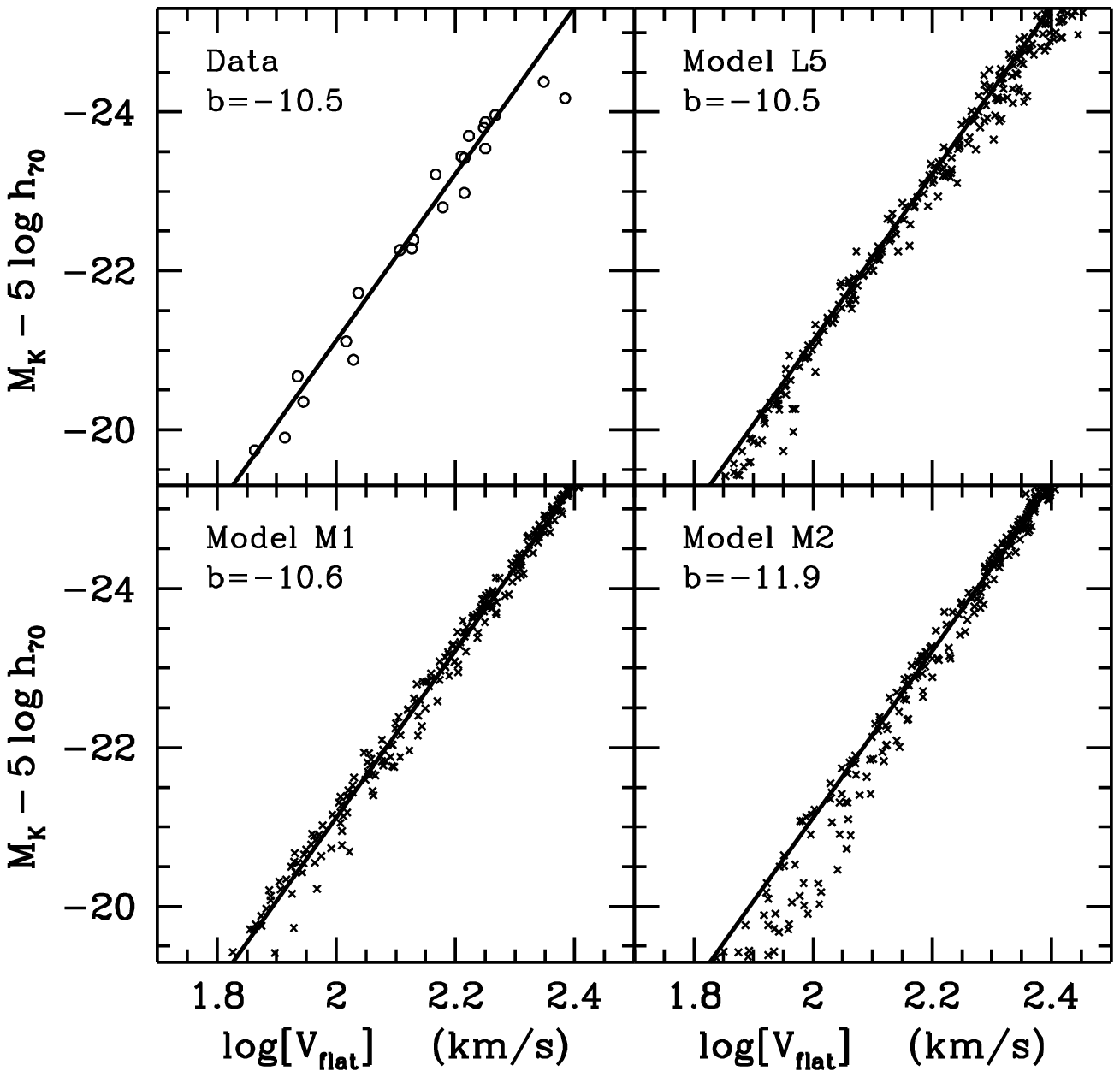}}
\ifsubmode
\vskip3.0truecm
\setcounter{figure}{0}
\addtocounter{figure}{1}
\centerline{Figure~\thefigure}
\else\figcaption{\figcaptf}\fi
\end{figure}


\clearpage
\begin{figure}
\epsfxsize=14.0truecm
\centerline{\epsfbox{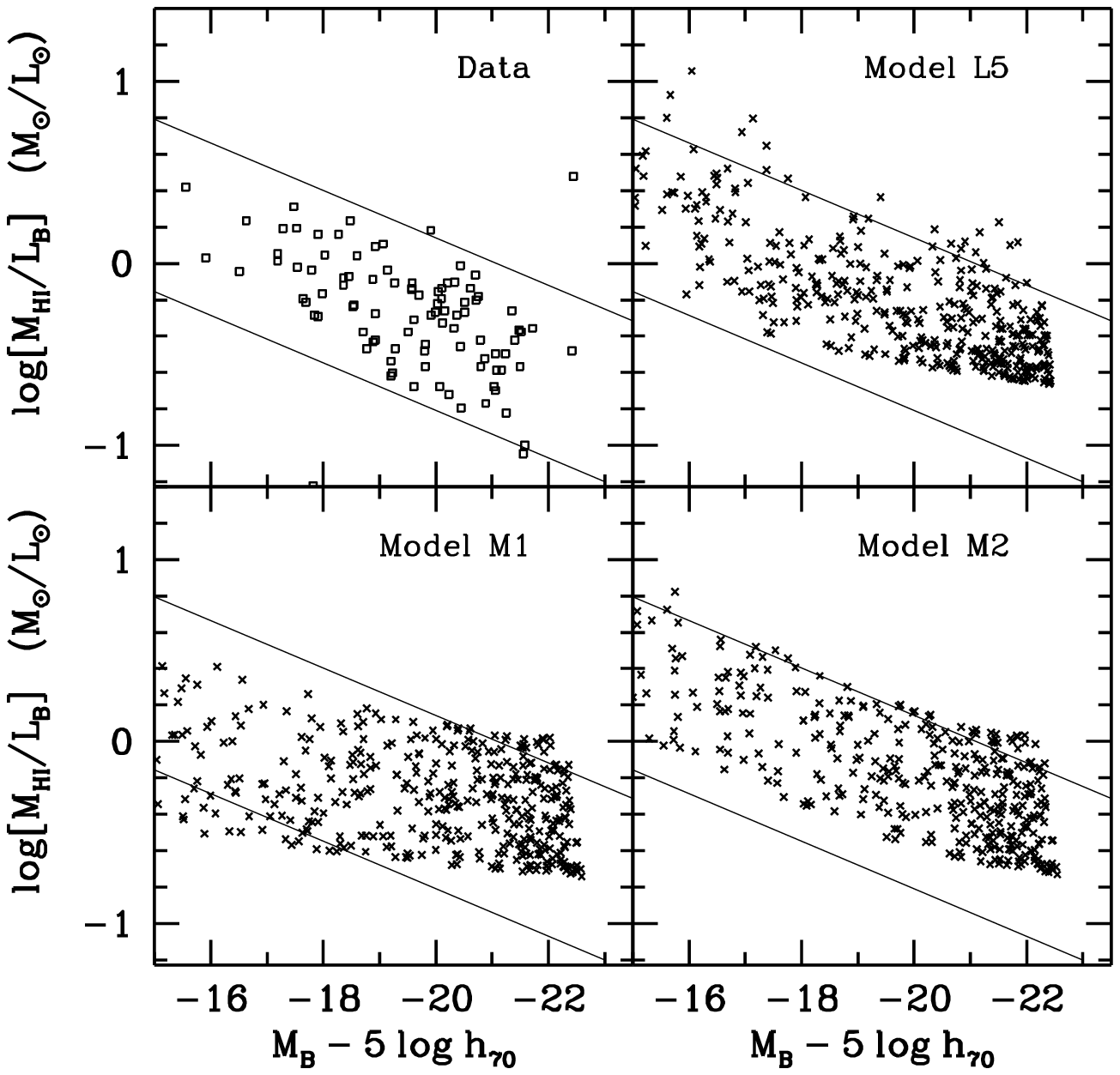}}
\ifsubmode
\vskip3.0truecm
\addtocounter{figure}{1}
\centerline{Figure~\thefigure}
\else\figcaption{\figcapgasmag}\fi
\end{figure}


\clearpage
\begin{figure}
\epsfxsize=14.0truecm
\centerline{\epsfbox{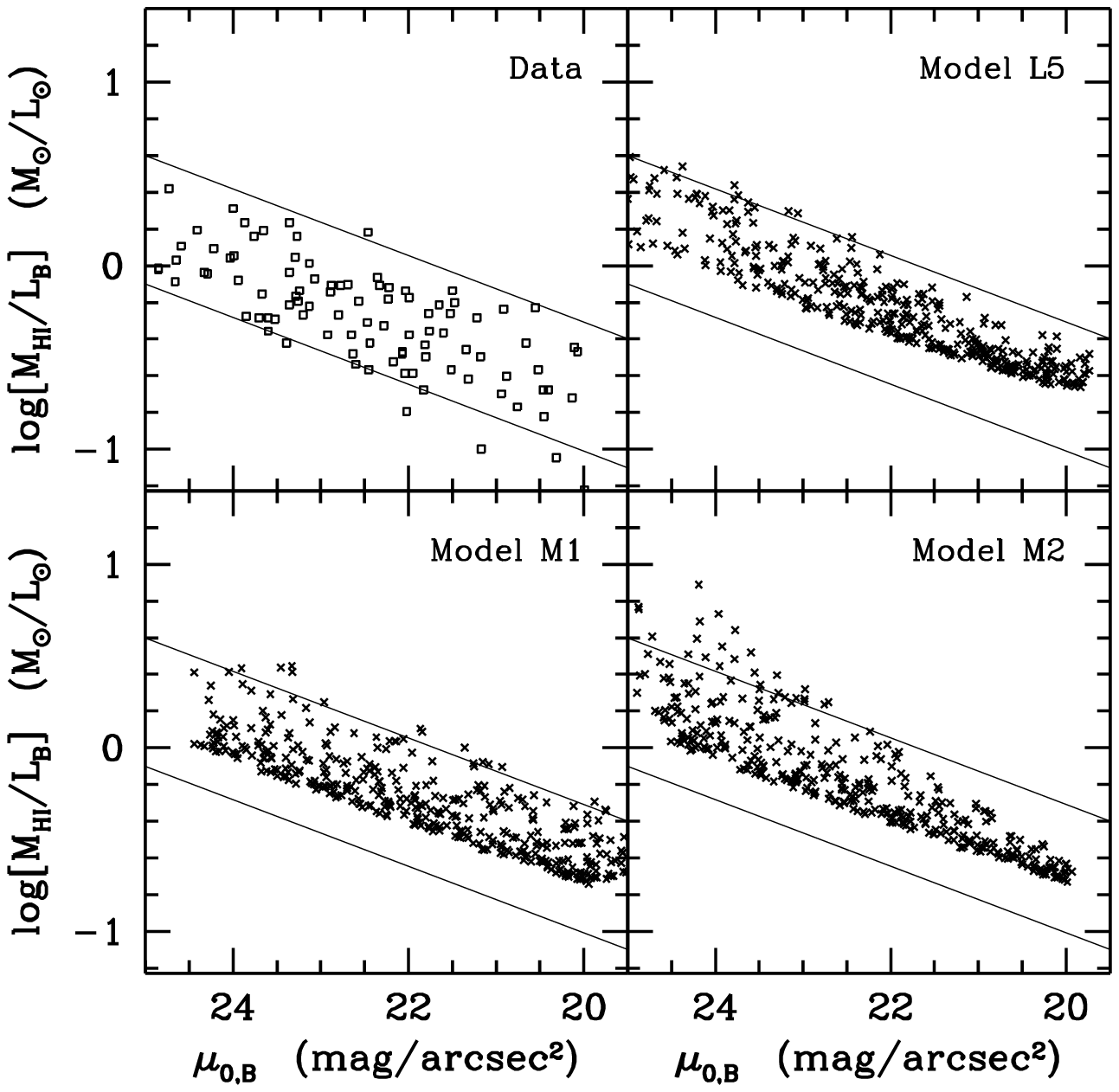}}
\ifsubmode
\vskip3.0truecm
\addtocounter{figure}{1}
\centerline{Figure~\thefigure}
\else\figcaption{\figcapgassb}\fi
\end{figure}


\clearpage
\begin{figure}
\epsfxsize=14.0truecm
\centerline{\epsfbox{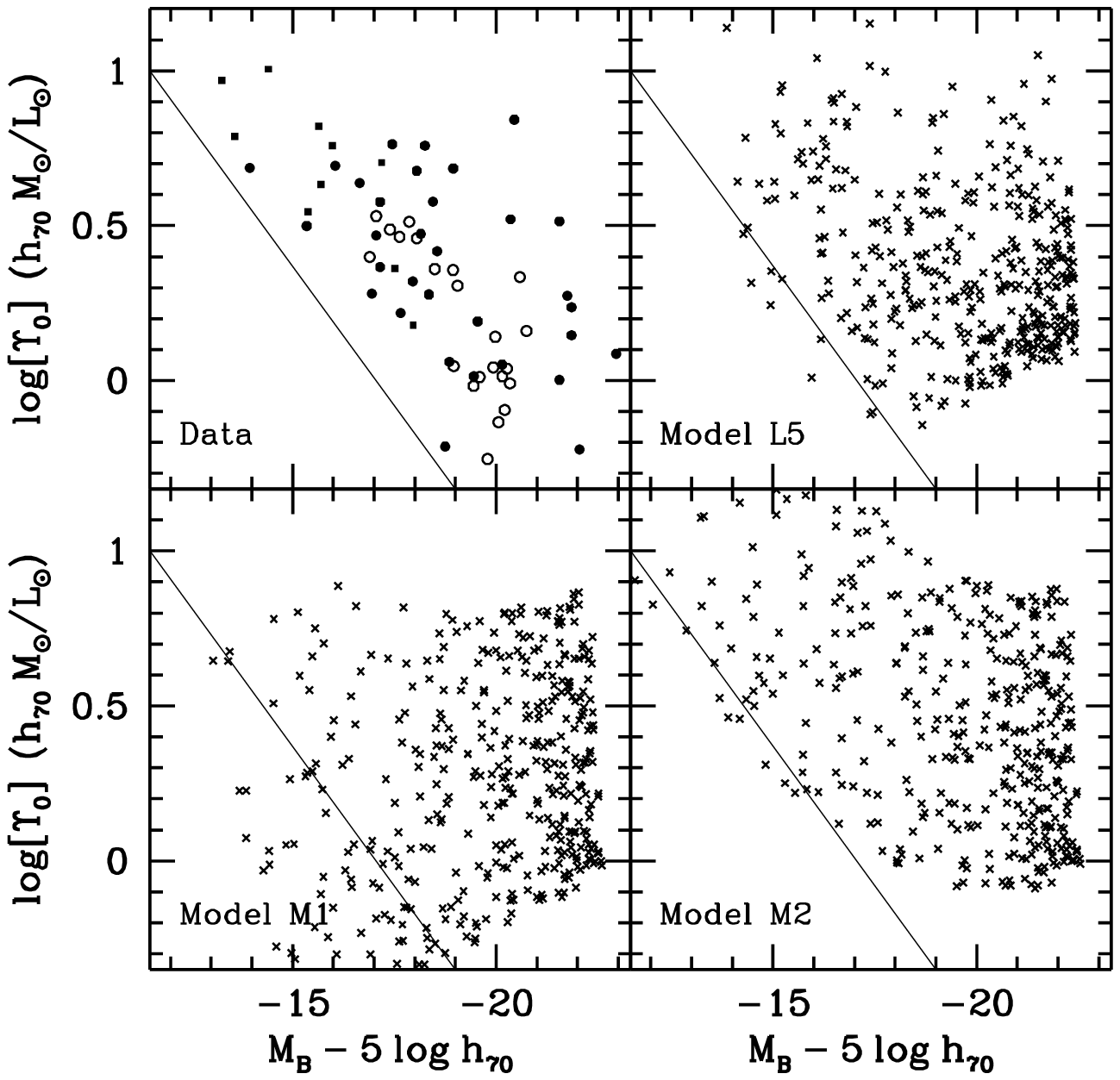}}
\ifsubmode
\vskip3.0truecm
\addtocounter{figure}{1}
\centerline{Figure~\thefigure}
\else\figcaption{\figcapupsmag}\fi
\end{figure}


\clearpage
\begin{figure}
\epsfxsize=14.0truecm
\centerline{\epsfbox{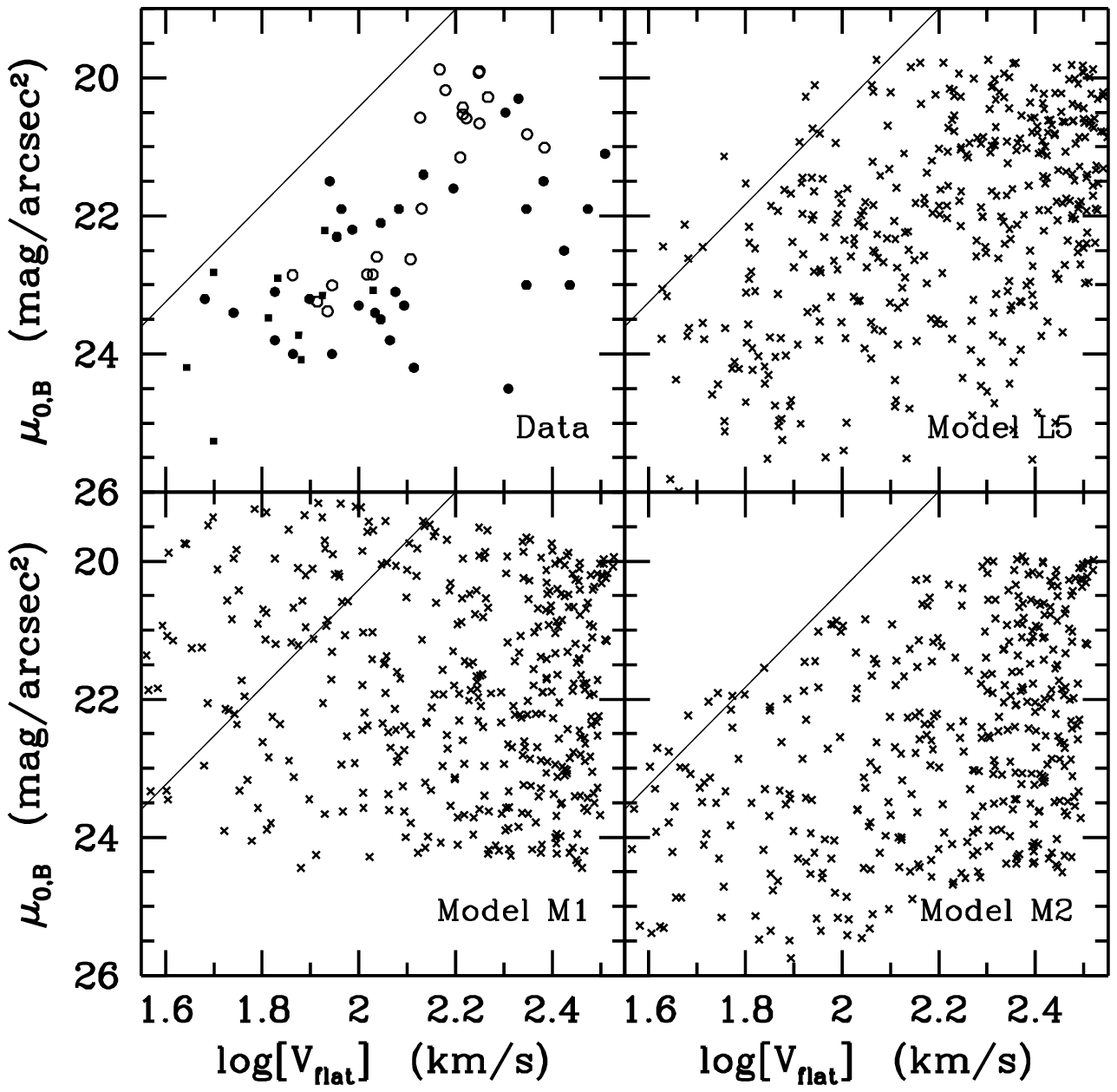}}
\ifsubmode
\vskip3.0truecm
\addtocounter{figure}{1}
\centerline{Figure~\thefigure}
\else\figcaption{\figcapsbvel}\fi
\end{figure}


\clearpage
\begin{figure}
\epsfxsize=14.0truecm
\centerline{\epsfbox{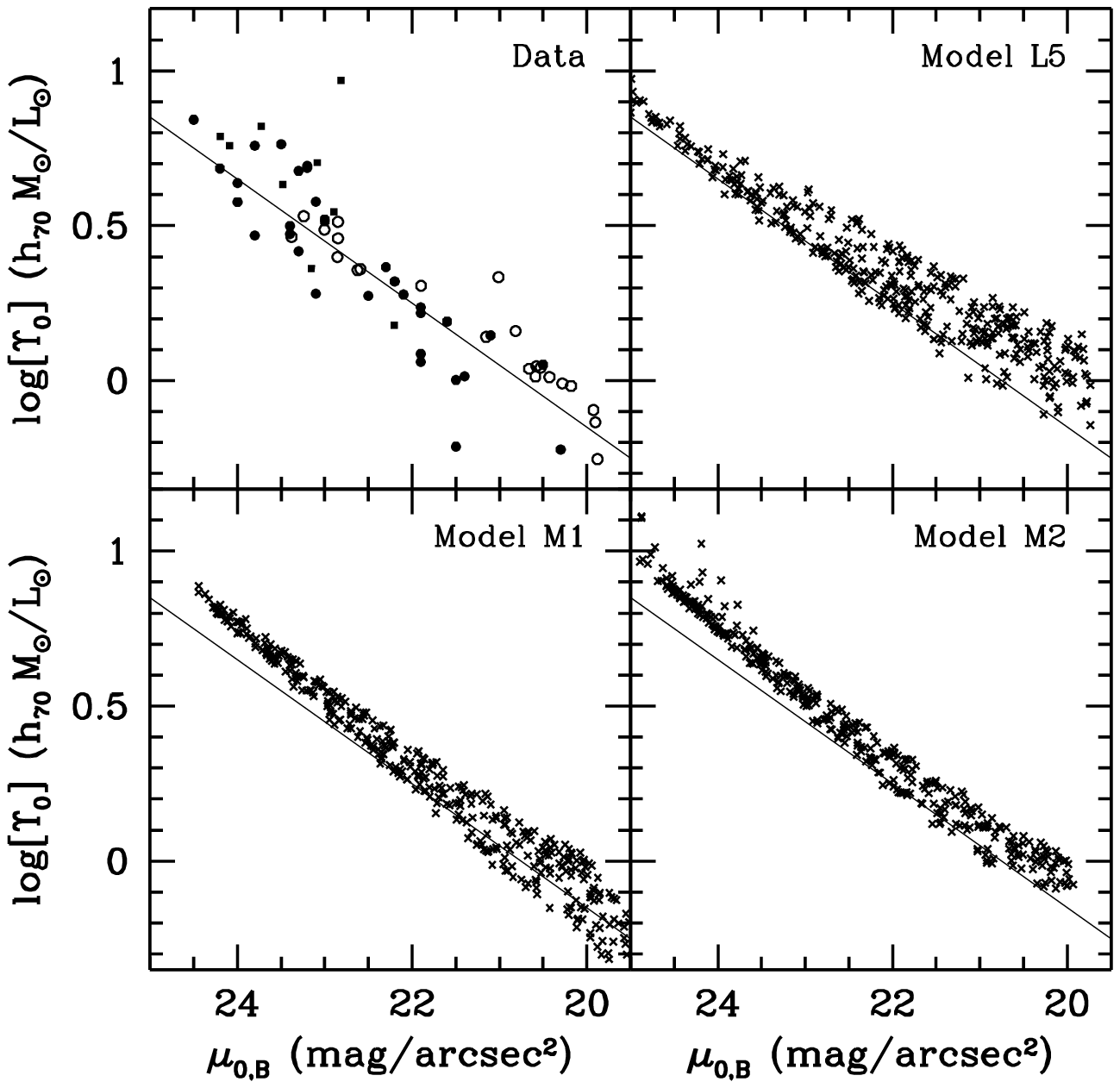}}
\ifsubmode
\vskip3.0truecm
\addtocounter{figure}{1}
\centerline{Figure~\thefigure}
\else\figcaption{\figcapupssb}\fi
\end{figure}


\clearpage
\begin{figure}
\epsfxsize=14.0truecm
\centerline{\epsfbox{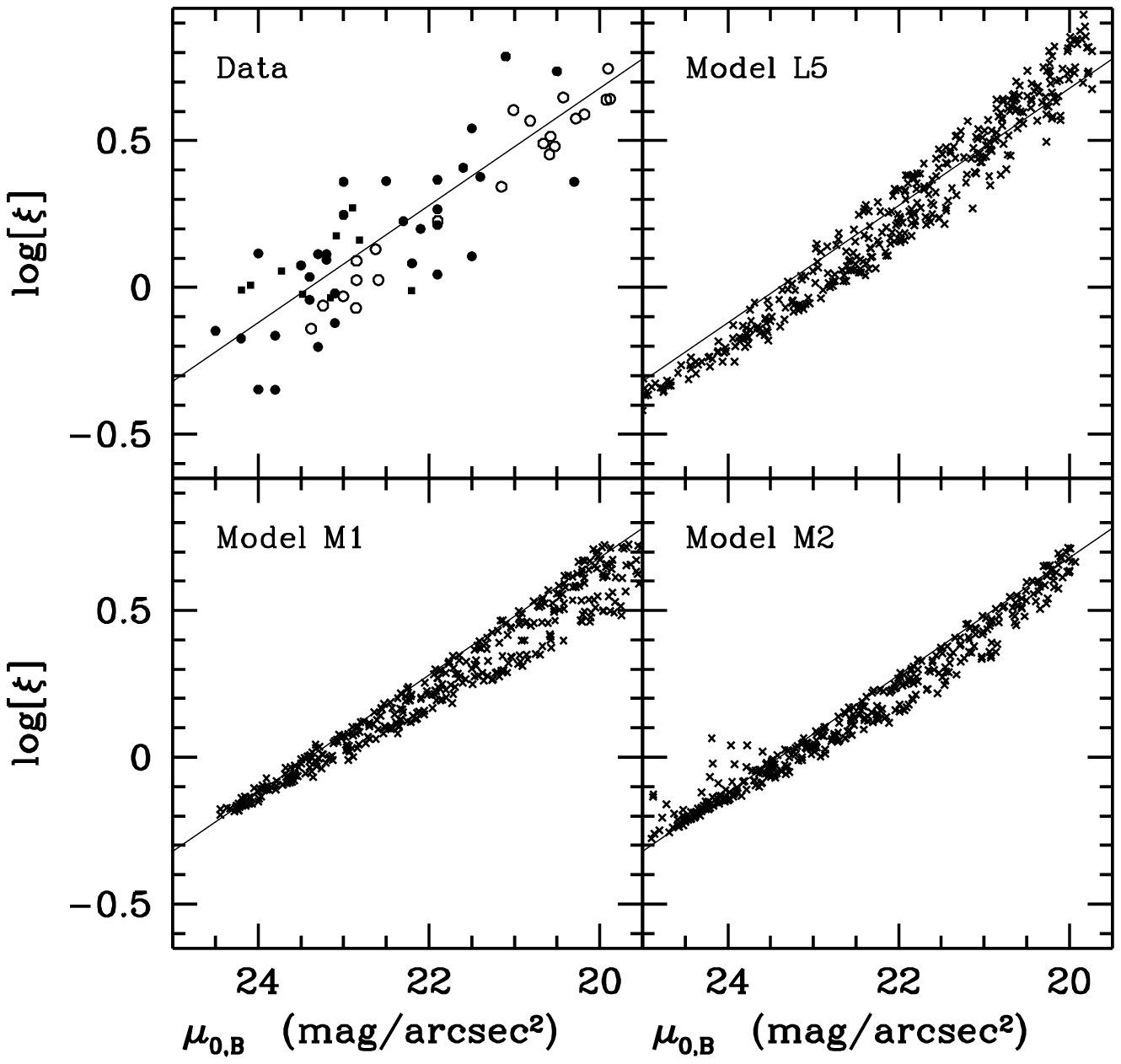}}
\ifsubmode
\vskip3.0truecm
\addtocounter{figure}{1}
\centerline{Figure~\thefigure}
\else\figcaption{\figcapxi}\fi
\end{figure}


\begin{figure}
\epsfxsize=16.0truecm
\centerline{\epsfbox{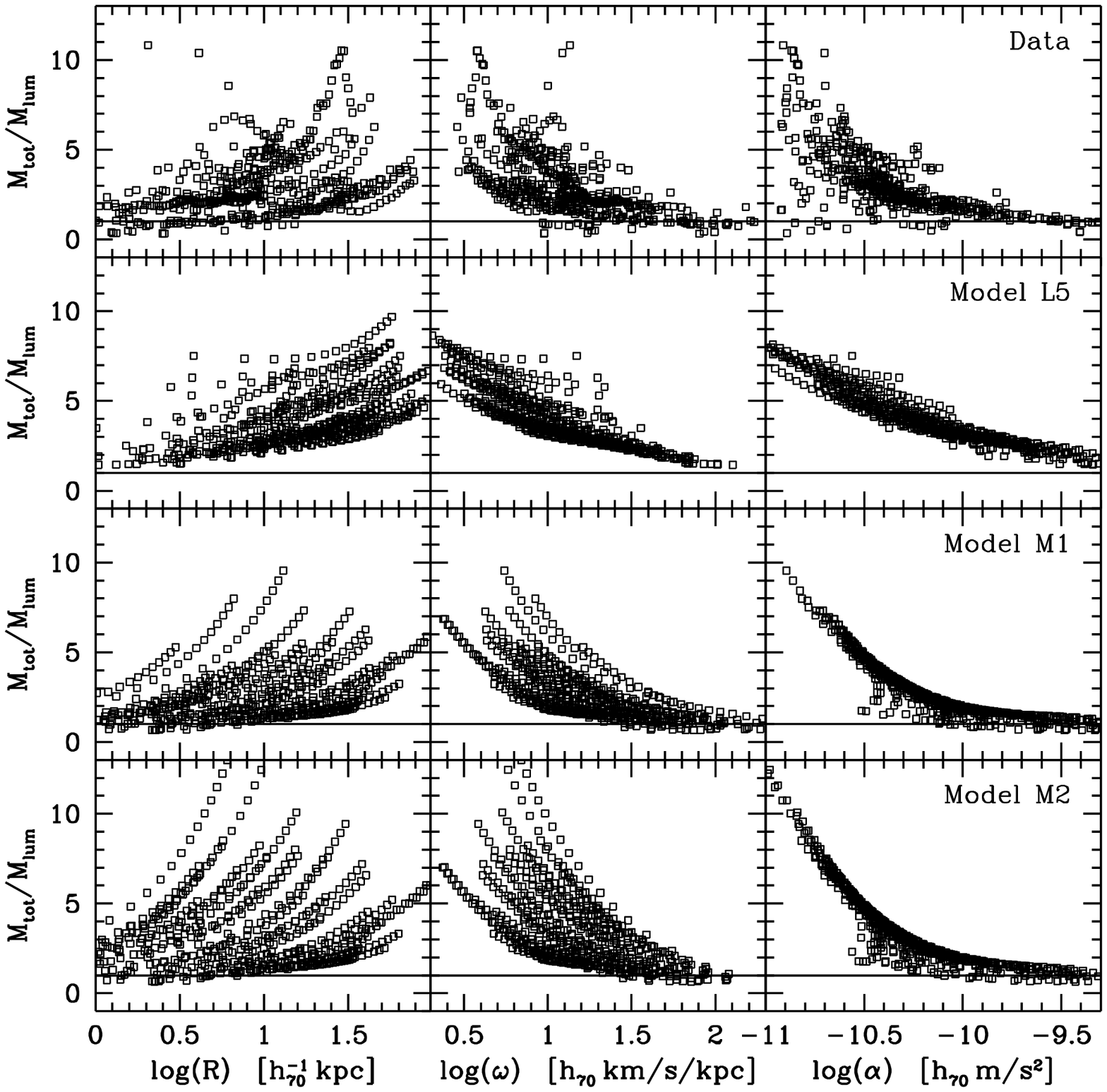}}
\ifsubmode
\vskip3.0truecm
\addtocounter{figure}{1}
\centerline{Figure~\thefigure}
\else\figcaption{\figcapacc}\fi
\end{figure}


\begin{figure}
\epsfxsize=16.0truecm
\centerline{\epsfbox{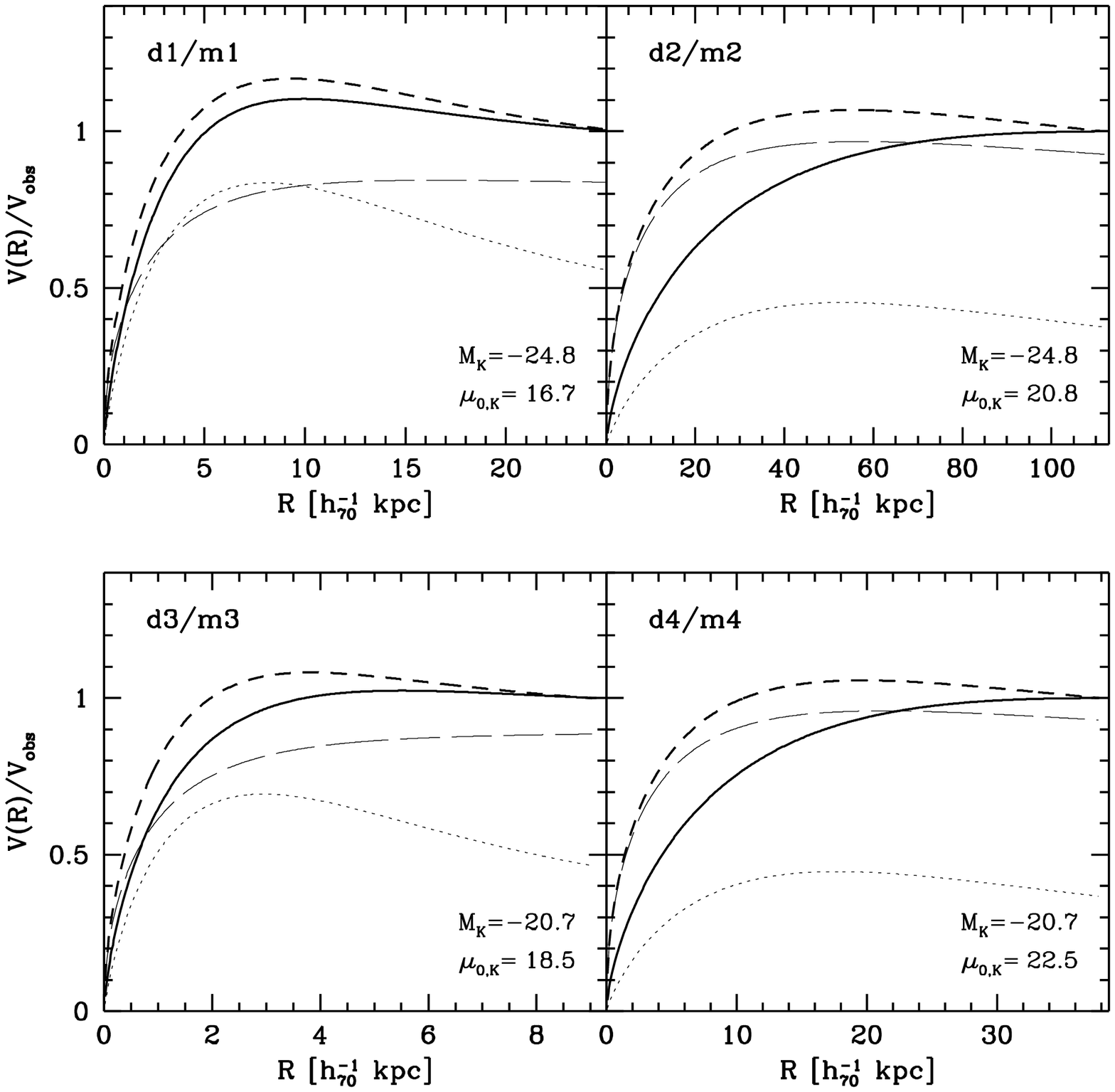}}
\ifsubmode
\vskip3.0truecm
\addtocounter{figure}{1}
\centerline{Figure~\thefigure}
\else\figcaption{\figcaprc}\fi
\end{figure}


\begin{figure}
\epsfxsize=17.0truecm
\centerline{\epsfbox{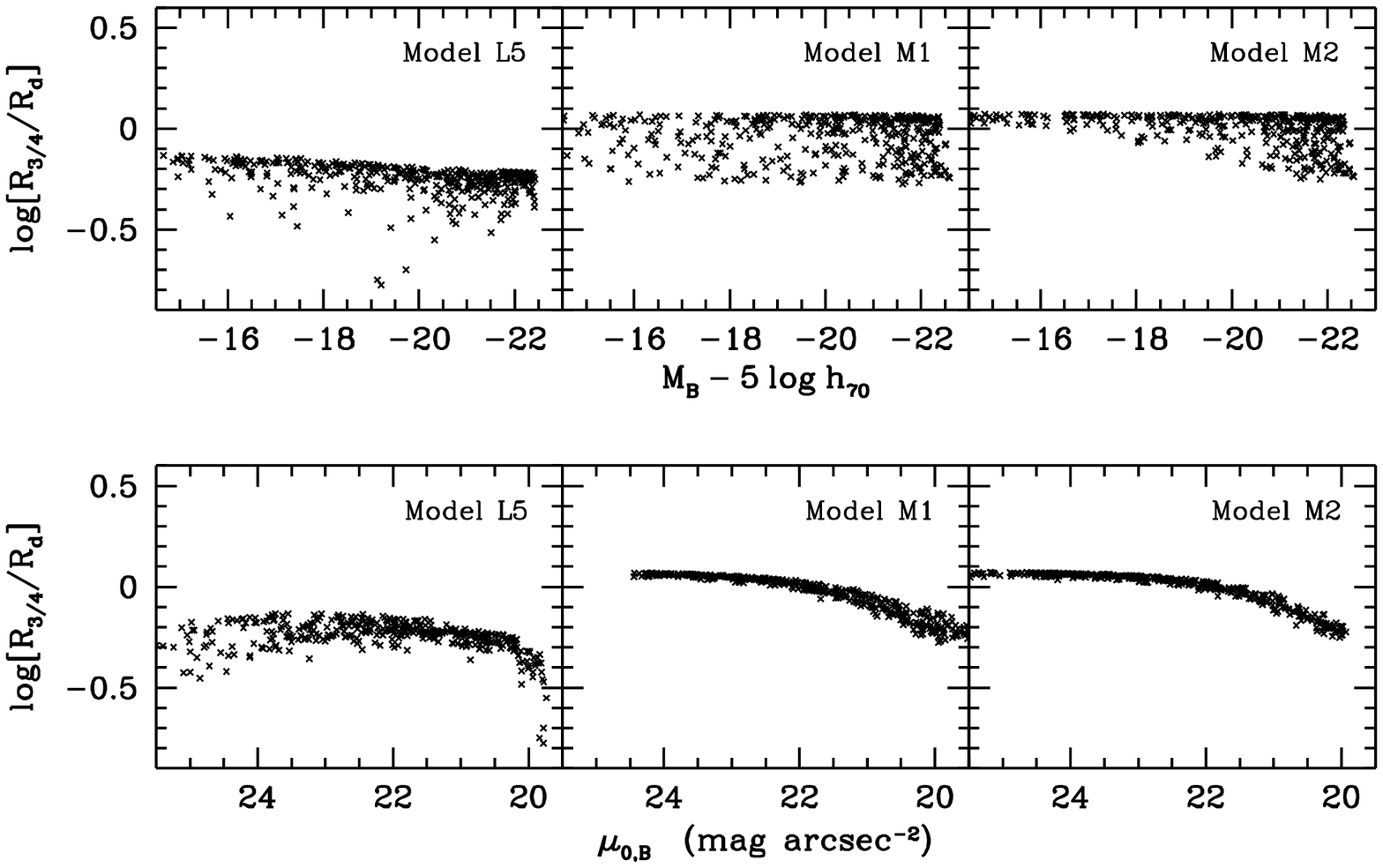}}
\ifsubmode
\vskip3.0truecm
\addtocounter{figure}{1}
\centerline{Figure~\thefigure}
\else\figcaption{\figcaprcrad}\fi
\end{figure}


\begin{figure}
\epsfxsize=16.0truecm
\centerline{\epsfbox{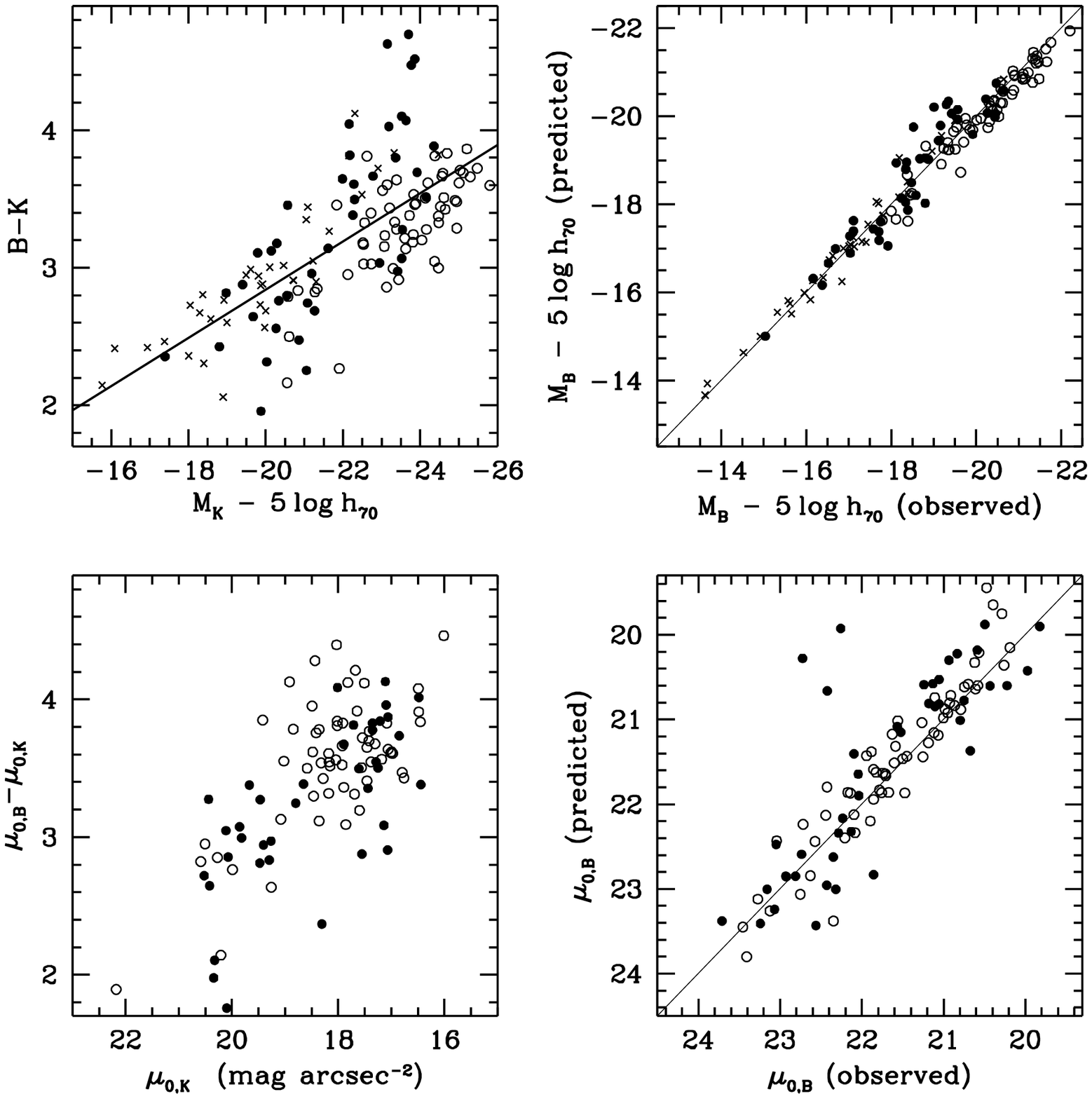}}
\ifsubmode
\vskip3.0truecm
\addtocounter{figure}{1}
\centerline{Figure~\thefigure}
\else\figcaption{\figcapcolors}\fi
\end{figure}


\fi


\clearpage
\ifsubmode\pagestyle{empty}\fi


\begin{deluxetable}{llcclrcc}
\tablecaption{Parameters of models discussed in the text. \label{tab:param}}
\tablehead{
\colhead{ID.} & \colhead{Hypothesis} & \colhead{$Q$} & 
\colhead{$\Upsilon^{*}_K$} &
\colhead{$\varepsilon_{\rm SN}^0$} & \colhead{$\nu$} &
\colhead{$b$} &\colhead{$\sigma_M$} \\ 
\colhead{(1)} & \colhead{(2)} & \colhead{(3)} & 
\colhead{(4)} & \colhead{(5)} & \colhead{(6)} &
\colhead{(7)} & \colhead{(8)} \\
}
\startdata
L5 & CDM  & $1.5$ & $0.40$ & $0.05$ & $-0.3$ & $-10.5$ & $0.20$ \\
M1 & MOND & $1.5$ & $0.53$ & $0.0$  &  $0.0$ & $-10.6$ & $0.21$ \\
M2 & MOND & $1.5$ & $0.53$ & $0.05$ & $-3.0$ & $-11.9$ & $0.33$ \\
\enddata

\tablecomments{Column~(1) lists the model ID,  by which we refer to it
  in the text.  Column~(2) indicates whether the model is based on the
  DM or the  MOND hypothesis.  Columns~(3) and~(4)  list  the value of
  the  Toomre parameter, $Q$,   and the stellar $K$-band mass-to-light
  ratio  in  $h_{70} \, \Msun/\Lsun$.     Columns~(5) and~(6) list the
  feedback     parameters   $\varepsilon_{\rm   SN}^0$    and   $\nu$,
  respectively.  Finally, columns~(7)  and~(8) list the slope $b$  and
  scatter  $\sigma_M$ (in  mag) of  the best  fitting  TF relation for
  model galaxies with $-19 \geq M_K - 5 {\rm log} h \geq -24$}
\end{deluxetable}

\clearpage


\begin{deluxetable}{cccrcrccc}
\tablecaption{Parameters of model galaxies discussed in the text. 
\label{tab:rcgal}}
\tablehead{
\colhead{ID.} & 
\colhead{Model} & 
\colhead{$M_K$} & 
\colhead{$R_d$} &
\colhead{$\mu_{0,K}$} &
\colhead{$V_{\rm flat}$} &
\colhead{$R_{3/4}/R_d$} & 
\colhead{$\lambda$} &
\colhead{$c$} \\
\colhead{} & 
\colhead{} & 
\colhead{mag} & 
\colhead{$h_{70}^{-1}$ kpc} &
\colhead{mag arcsec$^{-2}$} &
\colhead{$\kms$} &
\colhead{} & 
\colhead{} &
\colhead{} \\
\colhead{(1)} & \colhead{(2)} & \colhead{(3)} & 
\colhead{(4)} & \colhead{(5)} & \colhead{(6)} &
\colhead{(7)} & \colhead{(8)} & \colhead{(9)} \\
}
\startdata
d1 & L5 & $-24.8$ &  $3.8$ & $16.7$ & $237.5$ & $0.52$ & $0.040$ & $8.5$ \\
m1 & M1 & $-24.8$ &  $3.8$ & $16.7$ & $216.2$ & $0.68$ & $--$    & $--$ \\
   &    &         &        &        &         &        &         &      \\
d2 & L5 & $-24.8$ & $24.7$ & $20.8$ & $213.9$ & $0.41$ & $0.179$ & $8.3$ \\
m2 & M1 & $-24.8$ & $25.0$ & $20.8$ & $233.8$ & $1.17$ & $--$    & $--$ \\
   &    &         &        &        &         &        &         &      \\
d3 & L5 & $-20.7$ &  $1.4$ & $18.5$ &  $92.1$ & $0.63$ & $0.030$ & $9.4$ \\
m3 & M1 & $-20.7$ &  $1.4$ & $18.6$ &  $90.8$ & $0.98$ & $--$    & $--$ \\
   &    &         &        &        &         &        &         &      \\
d4 & L5 & $-20.7$ &  $8.3$ & $22.4$ &  $97.1$ & $0.46$ & $0.144$ & $9.8$ \\
m4 & M1 & $-20.7$ &  $8.3$ & $22.5$ & $114.9$ & $1.19$ & $--$    & $--$ \\
\enddata

\tablecomments{Column~(1) lists the model galaxy ID, by which we refer
  to it in the   text.  Column~(2) indicates   the model on  which the
  galaxy  is   based.  Columns~(3)-(5)  list the   $K$-band magnitude,
  $M_K$,  the  disk   scale-length, $R_d$,   and  the  central surface
  brightness,     $\mu_{0,K}$,      of      the    model     galaxies,
  respectively. Column~(6) lists the rotation velocities measured at a
  column  density of ${\rm N[HI]}  = 10^{20} {\rm cm}^{-2}$. The ratio
  of $R_{3/4}$ to $R_d$ is listed in column~(7).  Finally, columns~(8)
  and~(9) list  the spin  parameter  $\lambda$, and the  concentration
  parameter  $c$  of   the  NFW    halo   profile  (before   adiabatic
  contraction).  Since we only use these parameters  in the DM models,
  no values  are listed  for the   MOND  galaxies.  All values  listed
  correspond to $h_{70} = 1$}
\end{deluxetable}

\clearpage


\begin{deluxetable}{cccccc}
\tablecaption{Results. \label{tab:results}}
\tablehead{
\colhead{Observation} & 
\colhead{item} & 
\colhead{figure} &
\colhead{L5} & 
\colhead{M1} &
\colhead{M2} \\
\colhead{(1)} & \colhead{(2)} & \colhead{(3)} & 
\colhead{(4)} & \colhead{(5)} & \colhead{(6)} \\
}
\startdata
near-IR TF relation                  & 1 & 1    & $+$    & $+$  & $-$ \\
$M_{\rm HI}/L_B$                     & 4 & 2,3  & $+$    & $+$  & $+$ \\
$\Upsilon_0$ {\it vs.} $M_B$         & 3 & 4    & $+$    & $-$  & $+$ \\
$\Upsilon_0$ {\it vs.} $\mu_{0,B}$   & 3 & 6    & $+$    & $+$  & $+$ \\
$\mu_{0,B}$ {\it vs.} $V_{\rm flat}$ & - & 5    & $+$    & $-$  & $+$ \\
$\xi$ {\it vs.} $\mu_{0,B}$          & - & 7    & $+$    & $+$  & $+$ \\
$\Upsilon(R,\omega,\alpha)$          & 6 & 8    & $+/-$  & $+$  & $+$ \\
RC shapes                            & 2 & 9,10 &$+/-$?  & $+$? & $+$? \\
\enddata

\tablecomments{Column~(1)  indicates   the  observational constraint.  
  Columns~(2)   and~(3)   refer to the  item   number  of the  list of
  observational  facts  presented  in \S\ref{sec:intro},   and to  the
  number of  the figure in which a  comparison of the models  with the
  data is presented.  Columns~(4) to (6)  indicate whether each of the
  three  models  discussed   in this    paper  are consistent   ($+$),
  inconsistent ($-$), or  only marginally consistent  ($+/-$) with the
  data.  A  question mark  indicates  that more   data and/or  work is
  needed before a definite answer can be given.}
\end{deluxetable}

\clearpage



\clearpage

\begin{thebibliography}{}

\bibitem[]{Aar79}
Aaronson, M., Huchra, J., \& Mould, J. 1979, \apj , 229, 1

\bibitem[]{All79}
Allen, R. J., \& Shu, F. H. 1979, \apj , 227, 67

\bibitem[]{Ash92}
Ashman, K. M. 1992, \pasp , 104, 1109

\bibitem[]{BEf87} 
Barnes, J. E., \& Efstathiou, G. 1987, \apj , 319, 575

\bibitem[]{Beg91}
Begeman, K. C., Broeils, A. H., \& Sanders, R. H. 1991, \mnras , 249, 523

\bibitem[]{Bla98}
Blais-Ouelette, S., Carignan, C., \& Amram, P. 1998, preprint
  (astro-ph/9811142) 

\bibitem[]{Blu86} 
Blumenthal, G. R., Faber, S. M., Flores, R., \& Primack, J. R. 1986,
\apj , 301, 27

\bibitem[]{Bro92}
Broeils, A. H. 1992, Ph.D. Thesis, University of Groningen

\bibitem[]{Bul99}
Bullock, J. S., Kolatt, T. S., Sigad, Y., Somerville, R. S., Kravtsov,
A. V., Klypin, A. A., \& Dekel, A. 1990, preprint (astro-ph/9908159)

\bibitem[]{Bur95}
Burkert, A. 1995, \apj , 447, L25

\bibitem[]{Bur97}
Burkert, A., \& Silk, J. 1997, \apj , 488, L55

\bibitem[]{Bur82}
Burstein, D., Rubin, V. C., Thonnard, N., \& Ford, W. K. Jr. 1982,
  \apj , 253, 70

\bibitem[]{CCM89}
Cardelli, J. A., Clayton, G. C., \& Mathis, J. S. 1989, \apj , 345, 245
 
\bibitem[]{Cas91}
Casertano, S., \& van Gorkum, J. H. 1991, \aj , 101, 1231

\bibitem[]{Chr95} 
Christodoulou, D. M., Shlosman, I., \& Tohline, J. E. 1995, \apj , 443, 551

\bibitem[]{Cou97}
Courteau, S. 1997, \aj , 114, 2402

\bibitem[]{Dal97}
Dalcanton, J. J., Spergel, D. N., \& Summers, F. J. 1997, \apj , 482,
  659 (DSS97)

\bibitem[]{Dal99}
Dalcanton, J. J., et al. 1999, in preparation

\bibitem[]{deB96}
de Blok, W. J. G., McGaugh, S. S., \& van der Hulst, J. M. 1996,
  \mnras , 283, 18

\bibitem[]{BM98}
de Blok, W. J. G., \& McGaugh, S. S. 1998, \apj , 508, 132

\bibitem[]{deJ96a}
de Jong, R. S. 1996a, \aap , 313, 45

\bibitem[]{deJ96b}
de Jong, R. S. 1996b, \aaps , 118, 557

\bibitem[]{deJ94}
de Jong, R. S., \& van der Kruit, P. C. 1994, \aaps , 106, 451

\bibitem[]{Efs82} 
Efstathiou, G., Lake, G., \& Negroponte, J. 1982, \mnras , 19, 1069

\bibitem[]{Fal80}
Fall, S. M., \& Efstathiou, G. 1980, \mnras , 193, 189

\bibitem[]{Flo93} 
Flores, R., Primack, J. R., Blumenthal, G. R., \& Faber, S. M. 1993,
\apj , 412, 443

\bibitem[]{Flo94} 
Flores, R., \& Primack, J. R. 1994, \apj , 427, L1

\bibitem[]{For92}
Forbes, D. A. 1992, \aaps , 92, 583
 
\bibitem[]{Fre70}
Freeman, K. C. 1970, \apj , 160, 811

\bibitem[]{Fuk97}
Fukushige, T., \& Makino, J. 1997, \apj , 477, L9

\bibitem[]{Gav93}
Gavazzi, G. 1993, \apj , 419, 469

\bibitem[]{Gav96}
Gavazzi, G., Pierini, D., \& Boselli, A. 1996, \aap , 312, 397

\bibitem[]{Hof96}
Hoffman, G. L., Salpeter, E. E., Farhat, B., Roos, T., Williams, H.,
\& Helou, G. 1996, \apjs , 105, 269

\bibitem[]{Ken87}
Kent, S. M. 1987, \aj , 93, 816

\bibitem[]{Ken89}
Kennicutt, R. C. Jr. 1989, \apj , 344, 685

\bibitem[]{Ken98}
Kennicutt, R. C. Jr. 1998, \apj , 498, 541

\bibitem[]{Kly99}
Klypin, A. A., Kravtsov, A. V., Valenzuela, O., \& Prada, F. 1999,
  preprint (astro-ph/9901240)

\bibitem[]{Kor90}
Kormendy, J. 1990, in ASP Conf. Ser. No. 10, Evolution of the Universe
  of Galaxies, ed. R. G. Kron (Provo, UT, Brigham Young University
  Printing Services), 109 

\bibitem[]{Kra98}
Kravtsov, A. V., Klypin, A. A., Bullock, J. S., \& Primack,
  J. R. 1998, \apj , 502, 48

\bibitem[]{Lak89}
Lake, G. 1989, \apj , 345, L17

\bibitem[]{Lib92}
Liboff, R. L. 1992, \apj , 397, L71

\bibitem[]{Man89}
Mannheim, P. D., \& Kazanas, D. 1989, \apj , 342, 635

\bibitem[]{McG98}
McGaugh, S. S. 1998, preprint (astro-ph/9812327)

\bibitem[]{McG95}
McGaugh, S. S., Bothun, G. D., \& Schombert, J. M. 1995, \aj , 110, 573

\bibitem[]{MD97}
McGaugh, S. S., \& de Blok, W. J. G. 1997, \apj , 481, 689

\bibitem[]{MD98a}
McGaugh, S. S., \& de Blok, W. J. G. 1998a, \apj , 499, 41 (MB98a)

\bibitem[]{MD98b}
McGaugh, S. S., \& de Blok, W. J. G. 1998b, \apj , 499, 66 (MB98b)

\bibitem[]{Mil83a}
Milgrom, M. 1983a, \apj , 270, 365

\bibitem[]{Mil83b}
Milgrom, M. 1983b, \apj , 270, 371

\bibitem[]{Mil88}
Milgrom, M. 1988, \apj , 333, 689

\bibitem[]{Mil89}
Milgrom, M. 1989, \apj , 338, 121

\bibitem[]{Mil91}
Milgrom, M. 1991, \apj , 367, 490

\bibitem[]{Mo98}
Mo, H. J., Mao, S., \& White, S. D. M. 1998, \mnras , 295, 319

\bibitem[]{Mof96}
Moffat, J. W., \& Sokolov, I. Yu. 1996, Phys. Lett. B., 378, 59 

\bibitem[]{Moo94}
Moore, B., 1994, \nat , 370, 629

\bibitem[]{Moo98}
Moore, B., Governato, F., Quinn, T., Stadel, J., \& Lake, G. 1998,
\apj , 499, L5

\bibitem[]{Moo99a}
Moore, B., Ghigna, S., Governato, F., Lake, G., Quinn, T., Stadel, J.,
\& Tozzi, P. 1999a, \nat , in press
 
\bibitem[]{Moo99b}
Moore, B., Quinn, T., Governato, F., Stadel, J., \& Lake, G. 1999b,
  preprint (astro-ph/9903164)

\bibitem[]{Nav98} 
Navarro, J. F. 1998, preprint (astro-ph/9807084)

\bibitem[]{Nav96} 
Navarro, J. F., Frenk, C. S., \& White, S. D. M. 1996, \apj , 462, 563

\bibitem[]{Nav97} 
Navarro, J. F., Frenk, C. S., \& White, S. D. M. 1997, \apj , 490, 493

\bibitem[]{ODo94}
O'Donnell, J. E. 1994, \apj , 422, 1580

\bibitem[]{Per88}
Persic, M., \& Salucci, P. 1988, \mnras , 234, 131

\bibitem[]{Per90}
Persic, M., \& Salucci, P. 1990, \mnras , 245, 577

\bibitem[]{Per96}
Persic, M., Salucci, P., \& Stel, F. 1996, \mnras , 281, 27

\bibitem[]{Pic97}
Pickering, T. E., Impey, C. D., van Gorkum, J. H., \& Bothun, G. D.
  1997, \aj , 114, 1858

\bibitem[]{Pie88}
Pierce, M. J., \& Tully, R. B. 1988, \apj , 330, 579

\bibitem[]{Qui72}
Quirk, W. J. 1972, \apj , 176, L9

\bibitem[]{Rub80}
Rubin, V. C., Thonnard, N., \& Ford, W. K. Jr. 1980, \apj , 238, 471

\bibitem[]{Sal89}
Salucci, P., \& Frenk, C. S. 1989, \mnras , 237, 247

\bibitem[]{San99}
S\'anchez-Salcedo, F. J., \& Hildalgo-G\'amez, A. M. 1999, \aap , 345,
  36

\bibitem[]{San87}
Sancisi, R., \& van Albada, T. S. 1987, in Dark Matter in the
  Universe, IAU Symp. 117, eds. J. Kormendy \& G. R. Knapp (Reidel:
  Dordrecht), 67

\bibitem[]{San86}
Sanders, R. H. 1986, \mnras , 223, 539

\bibitem[]{San96}
Sanders, R. H. 1996, \apj , 473, 117

\bibitem[]{San98}
Sanders, R. H. \& Verheijen, M. A. W. 1998, \apj , 503, 97

\bibitem[]{Sch98}
Schlegel, D. J., Finkbeiner, D. P., \& Davis, M. 1998, \apj , 500, 525

\bibitem[]{Sch59}
Schmidt, M. 1959, \apj , 129, 243

\bibitem[]{Sco98}
Scorza, C., \& van den Bosch, F. C. 1998, \mnras , 300, 469

\bibitem[]{Spr95} 
Sprayberry, D., Bernstein, G. M., Impey, C. D., \& Bothun, G. D. 1995, \apj
, 438, 72

\bibitem[]{Sti99}
Stil, J. 1999, PhD. Thesis, Leiden University

\bibitem[]{Swa99}
Swaters, R. A. 1999, PhD. Thesis, University of Groningen

\bibitem[]{Sye99}
Syer, D., Mao, S., \& Mo, H. J. 1999, \mnras , 305, 357

\bibitem[]{Too64}
Toomre, A. 1964, \apj , 139, 1217

\bibitem[]{Tul77}
Tully, R. B., \& Fisher, J. R. 1977, \aap , 54, 661

\bibitem[]{Tul82}
Tully, R. B., Mould, J. \& Aaronson, M. 1982, \apj , 257, 527 

\bibitem[]{Tul96}
Tully, R. B., Verheijen, M. A. W., Pierce, M. J., Huang, J.-S., \&
Wainscoat, R. J. 1996, \aj , 112, 2471

\bibitem[]{Tul97}
Tully, R. B., \& Verheijen, M. A. W. 1997, \apj , 484, 145

\bibitem[]{vdB98}
van den Bosch, F. C. 1998, \apj , 507, 601

\bibitem[]{vdB99} 
van den Bosch, F. C. 1999, \apj , in press (astro-ph/9909501; paper I)

\bibitem[]{vdB99b}
van den Bosch, F. C., Robertson, B. E., Dalcanton, J. J., \& de Blok,
W. J. G.  1999, \aj , submitted (astro-ph/9911372)

\bibitem[]{vZ97}
van Zee, L., Haynes, M. P., Salzer, J. J., \& Broeils, A. H. 1997, 
  \aj , 113, 1618

\bibitem[]{Ver97}
Verheijen, M. A. W. 1997, PhD Thesis, University of Groningen

\bibitem[]{Vis81} 
Visvanathan, N. 1981, \aap , 100, L20

\bibitem[]{War92} 
Warren, M. S., Quinn, P. J., Salmon, J. K., \& Zurek, W. H. 1992, \apj
, 399, 405

\bibitem[]{Wys82}
Wyse, R. 1982, \mnras , 199, 1P

\bibitem[]{Whi97}
White, S. D. M. 1997, in Galaxy Scaling Relations: Origins,
  Evolution and Applications, eds. L. N. da Costa \& A. Renzini 
  (Springer-Verlag) 
 
\bibitem[]{Zwa95}
Zwaan, M. A., van der Hulst, J. M., de Blok, W. J. G., \& McGaugh,
S. S. 1995, \mnras , 273, L35

\end{thebibliography}
\end{document}